\documentclass[preprint,amsmath,amssymb]{revtex4}
\usepackage{graphicx}
\usepackage{dcolumn}
\usepackage{bm}
\begin{document}
\title{Hawking modes and the optimal disperser :\\Holographic lessons from the observer's causal-patch unitarity}
\author{${\rm J.\;Koohbor}^{\;(a)}$ \footnote {Electronic
address:~jkoohbor@ut.ac.ir}, ${\rm M.\;Nouri}$-${\rm Zonoz}^{\;(a)}$\footnote{Electronic
address:~nouri@ut.ac.ir, corresponding author} and ${\rm A. \;Tavanfar}^{\;(b)(c)}$ \footnote{Electronic
address:~alireza.tavanfar@cern.ch}}
\affiliation{(a): Department of Physics, University of Tehran, North Karegar Ave., Tehran 14395-547, Iran.\\
(b) : Instituto de Telecomunica\c{c}$\tilde{o}$es,
Physics of Information and Quantum Technologies Group, Lisbon, Portugal.\\
(c): International Institute of Physics, Universidade Federal do Rio Grande do Norte, 59078-400 Natal-RN, 
Brazil.}
\begin{abstract}
Based on an \emph{observer-centric methodology}, we pinpoint the basic origin of the spectral Planckianity of the asymptotic Hawking 
modes in the conventional treatments of the evaporating horizons.  
By considering an observer who analyzes a causal horizon in a generic spacetime, we first clarify how the asymptotic Planckian spectrum is imposed on the exponentially redshifted 
Hawking modes through a geometric dispersion mechanism developed by a semiclassical environment which is composed by all the modes that build up the curvature of 
the causal patch of the asymptotic observer. 
We also discuss the actual microscopic phenomenon of the Hawking evaporation of generic causal horizons. Our quantum description is based on a novel \emph{holographic 
scheme of gravitational open quantum systems} in which the degrees of freedom that build up the curvature of the observer's causal patch interact with the radiated 
Hawking modes, initially 
as environmental quanta, and after a crossover time, as quantum defects. Planckian dispersion of the modes would only be developed in the strict thermodynamic 
limit of this quantum environment, called 
\emph{optimal disperser}, which is nevertheless avoided holographically.  Finally, we outline and characterize how our microscopic formulation of the 
observer-centric holography, 
beyond the AdS/CFT examples and for generic causal patches, does realize the information-theoretic processing of unitarity. 
\end{abstract}
\maketitle

%-------------------------------------------------------------
\section{Motivation And Introduction} 
Quest for the universal and predictively successful theory of quantum gravity is still lively going on. The aim is to discover and formulate a complete theory of quantum gravity which is not only 
both self-consistent and non-perturbative in a completely and manifestly background-independent way, but also maximally generic in its phenomenological coverage. Through the long course of the collaborative research 
which have been conducted so far, a number of 
extremely deep lessons have been learned on \emph{how conceptually distinguished the correct formulation of quantum gravity theory must be}. 
Specially, the holographic trend through which gravity and (eventually 
the entire) spacetime are to emerge intrinsically is one most major lesson, with the AdS/CFT correspondence in String Theory being its 
currently best understood example \cite{HSBHR123M}. However 
a number of new major lessons are still expected to be learned, for us to be finally led to a complete and consistent set of principles 
forming the basis of a consistent, realistic theory of Quantum Gravity.\\
Up until now, in a void of direct or even helpfully indirect experimental data in quantum gravity, our advancement has mainly been guided by several methodologically significant theoretical observations and 
hints. Centered in the core of all these theoretical observations and hints is our present knowledge of \emph{quantum black hole physics} and 
the big challenge of discovering the correct and complete microscopic formulation of generic horizons. However, apart from the exotic and 
phenomenologically irrelevant cases of some supersymmetric black holes, a 
clear, consistent and correct understanding of \emph{the microscopic treatment of generic causal horizons} is yet missing.\\
We should highlight that, our incomplete understanding of the microscopic description of the generic black holes and their quantum behavior, is primarily not due to the lack of a  proper description of their 
near-singularity region, but instead from not yet having a correct understanding of the causal horizon itself, namely their corresponding \emph{holographic screen}. One good 
reason for this `priority discernment' is that the black hole physics, by means of the holographic principle, must be nothing but \emph{a decodification of a (non-gravitational) microscopic many-body physics
associated with the holographic screen}.\\
To obtain a consistent  microscopic description of a generic horizon, the first arena from which we need to clean up the inconsistencies of the semiclassical description, and replace them with the consistent 
microscopic physics, is the so-called \emph{information paradox}. Present study is taking some new steps to further advance in this same direction. Fortunately there are already very good 
reviews available on this subject \cite{30Rvs}, so that the background formulation of the paradox, and the repeating of the largely well-known common knowledge on this, could be escaped.\\ 
Now, let us give an outline of the present study.
In order to reinstate the indestructibility of the initial-state information in the Hawking evaporation of causal horizons, as the very first step, we must correctly pinpoint 
\emph{the most fundamental origin of their illusory 
semiclassical Planckian loss} which is common to the conventional treatments of black holes and other causal horizons. This \emph{root problem identification} 
is the main subject of \emph{the first part of this study}. To this end, we 
single out the original observation in \cite{MNZTP}, which was inspired by \cite{IP1SSP3}, and by a thorough re-analysis unfold its important messages. Let us highlight that 
the unique virtue of our study, in comparison with the most of the related literature, is its \emph{observer-based} approach.
First part of the paper includes sections two and three. In section two, we present our observer-centric analysis in the canonical example of Schwarzschild black hole 
(in arbitrary dimensions), and  verify its robustness and  generality by applying the same procedure to causal horizons in stationary spacetimes including the NUT hole and the Kerr black 
hole in section three.\\
By incorporating basically all the physically relevant deformations into our canonical setting, the same procedure is applied to other causal horizons 
with variations in geometry, topology  and asymptotics  of more spacetimes in sections four and five.\\
Second part of the paper, namely section six, which comes in three subsections, explores the microscopic restoration of the indestructibility of the quantum information. In the first subsection, 
we extract key microscopic 
lessons from the first part . In the second subsection, the physical and mathematical structure of a holographic microscopic formulation that realizes unitarity, is being developed. 
Finally in the third subsection, in order to set the generic theory of the evolving causal horizons, we interconnect that microscopic formalism 
with \emph{the fundamental observer-based theory} which will be developed in \cite{HUBMM}. We conclude in section seven.
%%%%%%%%%%%%%%%%%%%%%%%%%%%%%%%%%%%%%%%%%%%%%%%%%%%%%%%%%%%%%%%%%%%%%%%%%%%%
\section{Asymptotic Hawking Modes : The Canonical Example}
In this section, we define our \emph{observer-centric approach} and utilize it to analyze our canonical example, as the first case-study among 
a much broader family of spacetimes that will be explored later 
to confirm the generality and robustness of both this methodology and the conclusive statements we will learn from it. This canonical example sets 
up the basis of our novel understanding of the physical origin of 
the almost-Planckian spectrum of the asymptotic Hawking radiation. The \emph{information theoretic implications} of this understanding will be mainly 
addressed in \emph{second  part}. \\
We  base our study here upon the observationally-direct approach which was originally utilized in reference \cite{MNZTP}.  We take a most generic 
stance in our study, and explore example-wise the vast 
landscape of spacetimes which are solely characterized by the possession of a causal horizon, namely a (compact or non-compact) null hypersurface 
that connects as a boundary a complementary 
pair of causally disconnected sub-regions in the underlying spacetime. For every such spacetime, we  work out a thought experiment in which a 
sufficiently far freely falling  observer, named the famous Bob, receives 
and  spectrally analyzes an initially monochromatic light ray propagated from a freely falling source in the vicinity of the corresponding causal horizon. 
Using this same thought experiment, we will continue to explore a broad class of spacetimes with variations in the horizon geometry and 
topology,  as well as in the asymptotics.\\
Our canonical  case-study is  a static spherically symmetric spacetime possessing a quantum-evaporating compact event horizon, namely the 
Schwarzschild black hole. For the sake of 
generality, we will conduct the analysis of this  section in  arbitrary number of spacetime dimensions, but will be restricted to four spacetime 
dimensions in all our next examples. The metric 
of a one parameter family of Schwarzschild black holes in $D \geq 4$ spacetime dimensions is given by \cite{DWBB},
\begin{equation}  
ds^2=(1 -\dfrac{R_{H}^{D-3}}{r^{D-3}})dt^2 -(1 -\dfrac{R_{H}^{D-3}}{r^{D-3}})^{-1}dr^2-dS_{(D-2)}^2
\end{equation}
in which $dS_{D-2}$ is the metric of a $(D-2)$-dimensional sphere, and  
\begin{equation} 
R_{H} = (\frac{2Gm}{c^2(D-3)})^{\frac{1}{D-3}}  
\end{equation} 
is the radius of the spacetime event horizon.\\
Now let us realize our thought experiment in this geometry. Suppose that at some initial coordinate time  an \emph{initially monochromatic} null beam 
begins to propagate radially from a freely falling source at ${\cal P}({t_{in}, r_{in}})$  in the immediate vicinity of the horizon, 
say at $t=t_{in}, r_{in} = R_{H} + \epsilon$ with $\epsilon/R_{H} \ll 1$, and is eventually 
received and spectrally analyzed by the freely falling observer/receiver Bob at sufficiently large distance from the black hole at event ${\cal P}(t,r)$. 
It should be noted that neither Bob's position nor $\epsilon$ is fixed 
in Schwarzschild coordinates $(t,r)$. We also assume that Bob's location 
is far from the \emph{horizon's vacuum-entanglement zone} \cite{Bzone}. This is the horizon-exterior region with a width of the order of the 
horizon radius, which gives regional support 
to those of the Hawking modes which are almost-maximally entangled with their partner modes in the black hole's interior. \\
By the above assumptions,  take the Bob's receiving of the 
propagating  null rays to be identified with a late-time spacetime event ${\cal P}(t,r)$ with $r/R_{H} \gg 1$. The radial null trajectory is 
given by the null geodesics satisfying, 
\begin{equation} 
ds^2=(1 -\dfrac{R_{H}^{D-3}}{r^{D-3}})dt^2 -(1-\dfrac{R_{H}^{D-3}}{r^{D-3}})^{-1}dr^2=0  
\end{equation}
leading to the following equation 
\begin{equation} 
\dfrac{dr}{dt} =1 -\dfrac{R_{H}^{D-3}}{r^{D-3}} 
\end{equation}
for outgoing null rays. Solving this equation  \cite{GMR}, we end up with the following two different solutions depending on the 
spacetime dimension being odd or even,
\begin{equation}\label{t1}
D = {\rm odd:}\;\;\;\;t =r+\dfrac{R_{H}}{n} \ln \left( \dfrac{r-R_{H}}{r+R_{H}} \right)+{\dfrac{2R_{H}}{n}}\sum \limits_{k=1}^{n/2-1} P_k \cos {\dfrac{2k}{n}}\pi
-{\dfrac{2R_{H}}{n}}\sum \limits_{k=1}^{n/2-1} Q_k \sin {\dfrac{2k}{n}}\pi
\end{equation}
\begin{equation}\label{t2}
P_k = {\dfrac{1}{2}} \ln \left(x^2-2x \cos{\dfrac{2k}{n}\pi}+1\right),\;\;Q_k = \arctan\dfrac{x- \cos{\dfrac{2k}{n}\pi}}{\sin\dfrac{2k}{n}\pi}
\end{equation}
\begin{equation}\label{t3}
D = {\rm even:}\;\;\; t =r+\dfrac{R_{H}}{n} \ln \left(\dfrac{r-R_{H}}{R_{H}}\right)-{\dfrac{2R_{H}}{n}}\sum \limits_{k=0}^{(n-3)/2} P_k \cos {\dfrac{2k+1}{n}}\pi
-{\dfrac{2R_{H}}{n}}\sum \limits_{k=0}^{(n-3)/2} Q_k sin {\dfrac{2k+1}{n}}\pi
\end{equation}
\begin{equation}\label{t4}
P_k ={\dfrac{1}{2}} \ln \left(x^2+2x \cos{\dfrac{2k+1}{n}\pi}+1\right),\;\;Q_k = \arctan\dfrac{x+ \cos{\dfrac{2k+1}{n}\pi}}{\sin\dfrac{2k+1}{n}\pi}
\end{equation}
in which $x \equiv r/R_{H}$ and $n=D-3$. Now upon employing the initial conditions and taking into account that  $r \gg  R_{H}$ and $\epsilon \ll  R_{H}$, 
we find that to the leading order 
the trajectory is of the following form in both cases (up to an irrelevant constant term),
\begin{equation}\label{t5}
r \simeq t-t_{in} + {\dfrac{R_{H}}{D-3}} \;{\ln} \left(\frac{\epsilon}{R_{H}}\right).
\end{equation}
Now we use the fact that in a general spacetime for an emitter and an observer of null propagating modes (photons) on two different 
worldlines with 4-velocities $U^i_e$ and  $U^i_o$,
the relation between the  emitted and observed frequencies, at the two events ${\cal P}_e$ and ${\cal P}_o$, is given by \cite{Rindler},
\begin{equation}
\frac{\Omega_o}{\Omega_e} =  \frac{K_i({\cal P}_e) U^i({\cal P}_e)}{K_i({\cal P}_o) U^i({\cal P}_o)}
\end{equation}
where $K^i$ is the wave vector of the null beam  which is proportional to the tangent vector at each point on the corresponding null geodesic.
Upon restriction to the case of stationary spacetimes and in the {\it comoving} frames of the emitter and observer it reduces to \cite{Hobson},
\begin{equation}
\frac{\Omega_o}{\Omega_e} =  \left(\frac{g_{00}({\cal P}_e)}{g_{00}({\cal P}_o)}\right)^{1/2}
\end{equation}
Note the subtle point that this result is different from the redshift relation for {\it spatially fixed} emitter and receiver as stressed in 
the arguments of the $g_{00}$. This is due to the fact that there is an extra contribution to the redshift from the relative  velocity between the two observers \cite{Synge}. 
Now going back to our case, the relation for the redshift between the mode frequency 
$\Omega_{in}$ at the event ${\cal P}(r_{in},t_{in})$ and its frequency received at the event ${\cal P} (r,t)$  with $r \gg R_{H}$ 
(i.e where $g_{00}({\cal P}_o) \approx 1$), reduces to,
\begin{equation}
\Omega ({\cal P}(t,r))\; \simeq \; \Omega_{in}\;[g_{00}({\cal P}(r_{in},t_{in}))]^{1/2} 
\end{equation}
which, as a consequence of \emph{the presence of the corresponding causal horizon}, leads to an exponentially-redshifted frequency of the following form,
\begin{equation}\label{t6}
\Omega (t,r) \simeq \Omega_{in}\left[1-\left(\dfrac{R_{H}}{r_{in}}\right)^{D-3}\right]^{1/2} 
\simeq \Omega_{in} \left[(D-3)\dfrac{\epsilon}{R_{H}}\right]^{1/2} \simeq \Omega_{in}(D-3)^{\frac{1}{2}}
\exp\left[{\dfrac{-(D-3)(t-t_{in}-r)}{2R_{H}}}\right]
\end{equation}
in which in the last equation we substituted for  
$\epsilon (t)$ from equation (\ref{t5}), using the fact that for a radial null geodesic, the radial coordinate 
$r$ is an affine parameter which is at the same time proportional to the proper time of the freely falling radial observer/emitter. Indeed one could show that near the horizon ($\epsilon \ll R_H$) both the radial null geodesic  and radial timelike geodesic lead to   $\epsilon \propto e^{-\frac{t}{R_H}}$  \cite{Blau,Inv}.  The above result, which is pivotal in our  treatment of the asymptotic Hawking modes, could also be obtained in terms of the locally inertial 
null coordinate in the vicinity of the future 
horizon and the retarded Eddington-Finkelstein null coordinate which is locally inertial at infinity (refer to Eq. (2.98) in section 2.4.1 of \cite{Fab}).
In other words, both the source and observer LIFs employ the null geodesic connecting their 
corresponding poles, to define {\it a null geodesic coordinate systems} around their poles \cite{Rindler}.\\
Now, as equation (\ref{t6}) manifests, \emph{although being in a static spacetime}, the frequency of the modes measured by Bob at some  $r\gg R_{H}$ 
does non-trivially vary with $t$. 
Bob will not see monochromatic modes, even though they were initially radiated-out as a monochromatic beam.\\
This nontrivial time-dependence of the mode frequencies observed by Bob is a very crucial point. Obviously starting from the null geodesic equation 
means that we are treating the radiation in the geometric optics limit  and this was allowed by 
the fact that the freely falling observer receives the light over a small region (in his LIF) of space and time interval around him. Therefore, as 
in flat space, this in turn allows 
for the introduction of the wave surface of constant phase (eikonal), whose normal represents the direction of propagation. But the above 
nontrivial dependence of the mode frequency on 
coordinate time shows that Bob can go beyond the geometric limit by taking a wave packet, 
\begin{equation} \Phi(t,r)\propto \exp(i\theta(t,r)) \end{equation} 
to be centered on this null ray such that the instantaneous frequency would be related to the phase by 
\begin{equation} \Omega = \frac{\partial \theta}{\partial t}  \end{equation}
Now integrating \eqref{t6} with respect to the coordinate time, the relevant wave mode is found to be, 
\begin{equation}
\Phi(t,r) \propto \exp\left[i\int^t dt'\; \Omega(t',r) \right] 
\propto \exp\left[{-\dfrac{2 R_{H}}{(D-3)^{1/2}}} \;i\Omega_{in}\; \exp(-\dfrac{(D-3)(t-t_{in}-r)}{2 R_{H}})\right].
\end{equation}
Since for the asymptotic observer at far distance from the horizon $\tau \simeq t$, he could use the coordinate time  $t$ to Fourier decompose these modes with respect to the frequency $\omega$ in the following form,
\begin{equation}
\Phi(t,r)={\dfrac{1}{2\pi}} \int^{\infty}_{-\infty} d\omega\; e^{-i\omega t} f(\omega)
\end{equation} 
with the inverse transformation
\begin{equation}
f(\omega)= \int^{\infty}_{-\infty} dt\; e^{i\omega t} \;\Phi(t,r) \propto
\int^{\infty}_0 dx\;x^{{\dfrac{-2 i\omega R_{H}}{D-3}} - 1}\; \exp(-{\dfrac{2 R_{H}}{D-3}}ix \Omega_{in}) 
\end{equation}
where $x= \exp \left[(D-3)(-t + t_{in} + r)/ 2 R_{H} \right]$. Rotating the contour in the above integral to the imaginary axis, i.e. $x \rightarrow y=ix$, we have,
\begin{equation}
f(\omega) \propto e^{-\pi \omega R_{H} /(D-3)}\;\int^{i\infty}_0 dY\;Y^{z-1} e^{-Y}
\end{equation}
in which, \begin{equation} z=-2 i \omega  R_{H}/(D-3) \;\;\;{\rm and}\;\;\;Y=- 2 R_{H}\Omega_{in}y/(D-3) \nonumber 
\end{equation}
Now using representations of the Gamma function and that $|\Gamma (ix)|^2 = \pi /(x \sinh \pi x)$, the associated frequency spectrum is found to be \cite{MNZTP},
\begin{equation}
|f(\omega)|^{-2} \propto \exp(\dfrac{4\pi R_{H}\;\omega}{D-3})-1.
\end{equation}
Including the explicit dependence on the light speed $c$, spectrum of the \emph{distantly} observed Hawking modes becomes,
\begin{equation}\label{fm1}
|f(\omega)|^2 \propto \left[\exp\;(\dfrac{\omega}{\omega_0})-1\right]^{-1} 
\end{equation}
in which the \emph{characteristic frequency},
\begin{equation}\label{fm2}
\omega_0 = \dfrac{c}{4\pi} \cdot \dfrac{D-3}{R_{H}}, 
\end{equation}
apart from its dependence on the horizon scale and the dimension of the spacetime, is otherwise universal, as it should be.
What the observer detects, as read by the equation (\ref{fm1}), is obviously a Planckian spectrum, although the 
whole procedure we followed  was a  classical free field propagation in which no $\hbar$-dependence was included.
Obviously the microscopic mechanism by which the Hawking modes are created and propagated does intrinsically belong to the realm of quantum physics, and as such, in a more complete treatment, one should also consider the  backscattering which modifies the spectrum by the so called grey-body factor \cite{Fab}.\\ 
From the information-theoretic point of view, 
\emph{the phenomenon of the information loss is clearly present in the classically obtained equation (\ref{fm1})}. That is, as far as the information 
is being accounted for, 
the asymptotically observed Planckian multi-chromaticity of the initially monochromatic Hawking modes, is nothing but the semiclassical information loss.\\
So, the source of the Hawking radiation is a quantum effect, 
but the procedure which disperses the initial-state information content of the system of the Hawking modes as they propagate through Bob's causal-patch 
is geometrical and classical. 
In fact as we will 
later see in detail, it does turn out that:\\ \emph{The Planckian mismatch of information between the initial and the asymptotic Hawking modes is due to a classical 
geometric mechanism which itself can only occur in the strict thermodynamic limit
of a specific geometry in Bob's causal-patch}. The exact identifications of this will be given in the second part of the paper.
Further, we ask if in the aforementioned optimum dispersion of the information there also comes \emph{an actual or effective thermalization}. Intuitively, 
this question is natural to be 
raised, knowing 
the fact that the horizon itself is a well-defined thermal system with a finite temperature, and then also knowing that a Planckian profile information 
loss is typical of a black body radiation. 
For any kind of 
thermalization to be in this system, as a simple combination of the dimensional analysis together with the universality of this phenomenon implies, 
the associated temperature of the 
asymptotic radiation 
has to be proportional to a universal constant of dimension $\hbar$, namely, \emph{proportional to the $\hbar$ itself}. Therefore, the associated 
temperature would unavoidably be quantum-sourced. 
This point is the second crucial fact which we will later take notice of in second  part, in order to unfold the actual 
information-theoretic physics underlying the Hawking radiation. 
To see the quantum-thermal character of Hawking radiation let us suppose that Bob, who has been a radio astronomer so far, just renews his 
device to a quantum-particle detector, to see if he detects 
a Planckian power-spectrum for a 
corresponding collection of quantum beeps. By doing so, his plot for the asymptotic power spectrum will indeed fits the formula found by 
re-expressing the spectrum (\ref{fm1}) in terms of 
the energy quanta $E=\hbar\omega$, namely \cite{TP},
\begin{equation}
|f(\omega)|^2 \propto \left[\exp(\dfrac{\hbar \omega}{\hbar \omega_0}) -1 \right]^{-1} \propto \left[\exp(\dfrac{E}{k_BT}) -1\right]^{-1}.
\end{equation}
Therefore, the power spectrum of the asymptotic Hawking modes would be the Planckian blackbody distribution at temperature (including the explicit 
dependences on the constants $c$ and $K_B$),
\begin{equation}\label{ts}
T \;= \dfrac{\hbar c}{4\pi {k_B}}\;\cdot\;\frac{D-3}{R_{H}}
\end{equation}
which is indeed the same result obtained in the literature by different methods \cite{9KW}, and for $D=4$ does reduce to the Hawking temperature 
for a 4-d Schwarzschild black hole \cite{MNZTP}. \\
Before moving ahead to the next sections, let us here bring to the reader's attention an indirect insightful remark. It was previously observed 
in \cite{IP1SSP3} that by Fourier decomposing the complex Minkowski plane wave of a massless scalar 
field, $\Phi(t,{\bf x}) \propto e^{-i({\bf k}.{\bf x} - \omega t)}$, with respect to the proper time of 
a uniformly accelerated observer, one is interestingly led to a similar Planckian spectrum as seen by the observer. Therefore on the basis 
of \emph{the equivalence principle}, which allows replacing an 
accelerated observer with constant proper acceleration $g$ in flat spacetime with a static observer in a gravitational field of the same 
characteristic strength, one should have intuitively 
expected to obtain the same classical result (\ref{fm1}). 
Indeed, in ordinary (c.g.s) units, it was shown that \cite{IP1SSP3},
\begin{equation}\label{paddyetal}
P(\Omega) \propto \frac{1}{g}\;[\exp(\frac{c \Omega}{g/2 \pi})-1]^{-1}
\end{equation}
where $g$ is the proper acceleration of the observer and $\Omega$ is the frequency of the  decomposed wave component measured by the same observer. 
The Planckian spectrum (\ref{paddyetal}) is also a 
classical result, with no `$\hbar$' included. Moreover, the very procedure that had led to it was also entirely classical.  But yet indeed, it does 
precisely resemble its very twin result in the Unruh 
effect which is nevertheless a quantum field theoretic effect. Now once more, by simply rewriting the frequency-spectrum formula (\ref{paddyetal}) 
in the form of an energy-spectrum, being expressed in terms 
of the massless quantum bits of energy one obtains,
\begin{equation}\label{at}
P(E) \propto \frac{1}{g}\;[\exp(\frac{cE}{\hbar g/2 \pi })-1]^{-1}
\end{equation}
which is again a Planckian blackbody distribution at temperature, 
\begin{equation} 
T = \frac{\hbar g}{2\pi c k_B} 
\end{equation} 
This result indeed resembles very well, as it should, the closely-related case of the Rindler observer which we will address later. As we know, 
Rindler spacetime is specially interesting as it simulates the near-horizon region of typical black holes, so by invoking 
again the equivalence principle in equating the above temperature with the Hawking temperature of a 4-d Schwarzschild black hole (\ref{ts}) (including the explicit 
dependences on the constants $c$ and $G$) we obtain 
\begin{equation} 
g= \frac{c^4 k_B}{4GM} 
\end{equation} 
which is nothing but the surface gravity of a Schwarzschild black hole, fulfilling our intuitive expectation.
Finally, as a fact that we will verify example-wise and will also conceptually discuss later in this paper, let us mention the fact 
that:\\ The presence of a causally-disconnecting null boundary, i.e  a causal horizon, is central to the Planckian dispersion of the information.
%%%%%%%%%%%%%%%%%%%%%%%%%%%%%%
\section{Asymptotic Hawking Modes In Stationary Spacetimes\\\hspace*{1 cm}:\;\;\;The NUT Hole and The Kerr Black Hole}
The observer-based thought experiment method by which we derived and analyzed the semiclassical asymptotic aspects of the Hawking radiation in the canonical case of
Schwarzschild black hole, is  intrinsically different from the standard approaches to this subject. Therefore, we need to confirm if, beyond merely being a computational
coincidence or a simplest-case matching, we can trust this approach as a general, robust method in investigating the physics of the Hawking radiation from the quantum-evolving horizons. 
Thus, we have to examine if this approach is applicable to the generic spacetimes possessing causal horizons. Of course, we would not work out every possible example individually. Neither we would establish 
this verification by proving a formal theorem. But instead, we will manifest the robustness and generality of this approach by verifying example-wise that it is indeed stable under major physical 
deformations applied both on the spacetime geometry and on the virtues of the causal horizons. To this end, we will select a number of broadly different examples which are marked by the 
physical distinctions in the local, global, and in the causal structure of the spacetime. In particular, we consider deformations on the characteristics such as the symmetries and asymptotics of spacetimes,
as well as their horizon topologies and the very nature of the causal horizons. It should also be highlighted that this verification is a necessity, primarily because our aim is to look for a consistent 
resolution of the black hole information paradox, and to develop the correct microscopic theory of the evolving horizons which will be discussed in the second  part of this study.
In this section, we examine our approach in obtaining the characteristics of the asymptotic Hawking radiation in the case of \emph{stationary spacetimes with event horizons} which include two 
characteristically different spacetimes, namely \emph{the NUT hole} and the \emph{Kerr black hole}. 
%%%%%%%%%%%%%%%%%%%%%%%%%%%%%%%%%%%%%%%%%%%%%
\subsection{The NUT Hole}
We begin with the NUT spacetime which is physically the most straightforward generalization of the Schwarzschild spacetime. Concerning the symmetries of its metric tensor, the NUT geometry is stationary 
and axially-symmetric. But, in fact, from the physical point of view the NUT hole is a spherically symmetric spacetime. To support this assertion, we first note that the physical symmetries of a spacetime 
are primarily defined by the transformations under which its physical gauge-invariant observables stay invariant. On the other hand since the spacetime metric itself is not directly a physical 
observable, its isometries 
should not necessarily coincide with the physical symmetries of the spacetime. The NUT hole realizes this distinction clearly, because (unlike its metric) all its curvature invariants are found to be spherically 
symmetric \cite{LBMNZ}. Secondly, in our observer-centric approach, one can identify the physical symmetries of the spacetime as the symmetries of the physical observables which are defined with respect to an input 
family of observers. Specifically, in the threading decomposition of spacetime, Einstein  field equations could be re-expressed in a quasi-Maxwell form in terms of the gravitoelectric and gravitomagnetic fields defined 
with respect to a congruence of observer worldlines. Then in NUT spacetime  besides the mass parameter \emph{$m$}, identified with the gravitoelectric charge, the spacetime can also be endowed with a 
gravitomagnetic monopole charge \emph{$\ell$}.\\
Within the threading formalism, one can  show that in NUT spacetime, all the physical observables, including the gravitoelectric and gravitomagnetic fields, are indeed spherically symmetric.
The metric of the NUT spacetime can be written in the following form \cite{LBMNZ}, 
\begin{equation}\label{NUT}
ds^2 =f(r)(dt - 2l \cos\theta d\phi)^2 - \frac{dr^2}{f(r)} - (r^2 +l^2) d\Omega^2\;\;\;;\;\;\;f(r) =\dfrac{r^2-2mr-l^2}{r^2+l^2}
\end{equation} 
The spacetime being physically spherical has radial null geodesics which are given by \cite{GPB},
\begin{equation}
\dfrac{dr}{dt} =\dfrac{r^2-2mr-l^2}{r^2+l^2}\equiv\dfrac{(r-r_+)(r-r_-)}{r^2 +l^2}
\end{equation} 
in which $r_\pm=m\pm(m^2+l^2)^{1/2}$ are the two event horizons of the NUT solution. The above equation integrated leads to,
\begin{equation}
t=r+r_+\ln\left| \dfrac{r}{r_+}-1\right|+r_-\ln\left| \dfrac{r}{r_-}-1\right|.
\end{equation}
which are the physically-anticipated radial null geodesic of the spacetime.
It should be highlighted that in the case of the NUT metric there are no intrinsic singularities hidden behind the horizon, justifying the given name NUT hole.
Also, it is to be noted that, the two horizons of 
the geometry, separate the stationary NUT regions ($r < r_-$ and $r > r_+$) from the time-dependent Taub region ($r_- < r < r_+$) of the so-called Taub-NUT 
spacetime which are causally disconnected \cite{GPB}. 
Now, let us apply and examine our approach to this causally-nontrivial spacetime. As set in the canonical example, we consider a beam of initially-monochromatic 
massless Hawking modes that propagate radially 
in the immediate proximity of the outer event horizon $(r_+ + \epsilon)$ at time $t=t_{in}$, to be received by the asymptotic 
observer Bob at the late-time event ${\cal P}(t,r)$, with
$r \gg  r_+$ and $\epsilon \ll  r_+$. The equation of such trajectory has the following form,
\begin{equation}
r \simeq t-t_{in}+r_+\ln \left(\dfrac{\epsilon}{r_+}\right)
\end{equation}
The redshifted frequency $\Omega$ will be related to the initial frequency $\Omega_{in}$ in at $r_{in}$ by
\begin{equation}
\Omega=\Omega_{in} \left[g_{00}(r=r_+ + \epsilon)\right]^{1/2} \simeq \Omega_{in}(\frac{\epsilon}{r_+})^{1/2} \simeq \Omega_{in}\exp \left(\dfrac{-t+t_{in}+r}{2r_+}\right).
\end{equation}
By applying the very same detailed procedure used in the Schwarzschild example now to NUT spacetime, we finally obtain the following power spectrum for a wave packet which has scattered off the outer event horizon 
of the NUT hole reaching the observer Bob \emph{at late times and at $r \gg r_+$},
\begin{equation}\label{nutfrequency}
|f(\omega)|^2 \varpropto \left[\exp \left( \dfrac{1}{4 \pi(m+(m^2+l^2)^{1/2})}\right)\omega -1\right]^{-1}
\end{equation}
The result (\ref{nutfrequency}) confirms the validity of our method for the NUT hole. Indeed, it again is the \emph{classically-obtained Planckian frequency spectrum} of the asymptotic Hawking modes, which being 
re-expressed in terms of the energy modes, leads to the Planckian radiation at the following temperature, 
\begin{equation}\label{nutt}
T\;=\;\frac{\hbar}{4\pi \left[ m+(m^2+l^2 )^{1/2}\right]}
\end{equation} 
which matches the standard result in the literature \cite{11HHP,12KM}.
%%%%%%%%%%%%%%%%%%%%%%%%%%%%%%%%%%%%%%%%%%%%%%%%%%%%5
\subsection{The Kerr Black Hole}
The second example is the Kerr black hole. The method has already been worked out in \cite{MNZTP}. Kerr black hole is a two-parameter stationary spacetime, which being axially 
symmetric both mathematically 
and physically, does not have radial null geodesics and so one should use principal null congruences to apply the Fourier decomposition method. It also contains two horizons 
which not only divide the spacetime 
into causally disconnected regions but also hide an intrinsic singularity from the outer region. Another point is the fact that, the infinite redshift surface and the event 
horizon do not coincide. This will 
require a careful employment of the above procedure taking into account the existence of the ergoregion from inside which  and in the vicinity of the outer horizon 
the outgoing null rays propagate to be received by observers at late time and large distances.
Working out all these details, we are led to the Planckian spectrum \cite{MNZTP},
\begin{equation}
|f(\omega)|^2 \varpropto \left[\exp \left( \dfrac{4 \pi(m^2+m(m^2 - a^2)^{1/2})}{(m^2 - a^2)^{1/2}}\right)\omega -1\right]^{-1}
\end{equation}
at the temperature, $$T\;=\;\hbar\;\dfrac{(m^2 - a^2)^{1/2}}{4 \pi(m^2+m(m^2 - a^2)^{1/2})}$$ which indeed reduces to the result for the Schwarzschild black hole for $a=0$.\\
Before taking up the task to discuss the question of unitarity of the information received by Bob in the second part of this study it should be noted that
as pointed out earlier, the above procedure in obtaining the Planckian spectrum of outgoing null rays propagated from the vicinity of 
causal horizons is a robust and general procedure. To show this robustness in a much broader setting 
which includes variations in geometry, topology and  asymptotics 
of the spacetimes as well as the inclusion of spacetimes with observer-based horizons, we will consider more examples in following sections.
%%%%%%%%%%%%%%%%%%%%%%%%%%%%%%%%%%%%%%%%%%%%%%%%%%%%%%%%%%%%%%%%%%%%%%%%%%%%%%%%%%%%
\section{Asymptotic Hawking Modes in a much broader setting : Variations of Geometry, Topology And Asymptotics}
In this section we examine our approach under other deformations in the physical and mathematical settings where Hawking radiation can be configured. 
In the first two of these threefold examples, we consider different global phases of the observer's causal-patch geometry, realized by the two possible deformations 
of the spacetime asymptotics while the horizon 
geometry remains intact. These two examples are the Schwarzschild black hole with the dS and the AdS asymptotics, respectively. 
In the third example, we consider the opposite setting in which the asymptoticss of the spacetime are freezed, but the horizon topology is  maximally deformed. By a maximal deformation of the horizon 
topology, we mean that its physically most relevant topological characteristic, namely the compactness of the causal screen, will be deformed. We will address
spacetimes with non-compact causal surfaces, namely, causal domain walls of infinite area, which are event horizons of finite temperature, and so radiate-out Hawking modes. The specific spacetime that we choose, 
is the simplest and best representative of this class of examples, namely the asymptotically-AdS cylindrical black hole whose solution and thermal behavior were initially obtained in \cite{L}.  
To make the presentation compact, let us begin with the general form of a class of metrics that collectively encodes the solutions of these three examples. This general metric, in terms of the mass parameter $m$ 
and the cosmological constant $\Lambda$, is given by \cite{V},
\begin{equation}\label{sch-ds}
ds^2=\left(k-\dfrac{2m}{r}-\dfrac{\Lambda}{3}r^2\right)dt^2-\left(k-\dfrac{2m}{r}-\dfrac{\Lambda}{3}r^2\right)^{-1}dr^2-r^2d\Omega_k^2
\end{equation}
The curvature parameter $k$ takes the values $\lbrace -1,0,+1\rbrace $, so that $d\Omega_{(k)}^2$ is the metric on a Riemann surface $\Sigma_{(k)}$ of constant Gaussian curvature $k$. The surface 
$\Sigma_{(k)}$, which can be of different topologies, will be the quantum-evaporating event horizon for the corresponding black hole solutions. When expressed in terms of the angular coordinates $(\theta,\phi)$, 
the area element $d\Omega_{(k)}^2$ takes the following forms,
\begin{equation}
d\Omega_{(k)}^2 = \begin{cases}
 \; d\theta^2 + \sin^2(\theta) d\phi^2, \;\;\;\;\;\;\; k=1\\ 
 \; d\theta^2 +  d\phi^2, \;\;\;\;\;\;\;\;\;\;\;\;\;\;\;\;\;\; k=0\\
 \; d\theta^2 + \sinh^2(\theta) d\phi^2, \;\;\;\; k=-1
\end{cases}\nonumber	
\end{equation}
The AdS cylindrical black hole corresponds to the above metric with $k=0$ and with a negative $\Lambda$, while both the Schwarzschild AdS and the Schwarzschild dS spacetimes are given by this same metric with 
negative and positive values of `$\Lambda$' respectively, upon taking the choice `$k=1$'.
%%%%%%%%%%%%%%%%%%%%%%%%%%%%%%%%%%%%%%%%%%%%%%%%%%
\subsection{Schwarzschild horizon in de Sitter Phase}
The de Sitter phase of quantum gravity (i.e $\Lambda > 0$), is drastically distinguished from the other two phases with $\Lambda \leq 0$, for two reasons. \emph{First}, the physical 
supersymmetric enhancements are not possible in the de Sitter phase. \emph{Second}, the de Sitter phase of quantum gravity should be holographically formulated by quantum theories which are strictly 
finite-dimensional \cite{BFA,TBA}. The theory of black holes in this phase is not only a physical sector of it, but also serves as a precious laboratory in developing the full theory of the 
\emph{de Sitter quantum gravity}. Here we employ our principal method to the de Sitter phase with the  motivation of formulating the unitarity of the horizon evolution in the second  part of this work.
Starting with the example of the Schwarzschild-dS horizon whose radial null geodesics are integrated from,
\begin{eqnarray}
\dfrac{dr}{dt}=1-\dfrac{2M}{r}-\dfrac{\Lambda r^2}{3},
\end{eqnarray}
we restrict our attention to the case $\Lambda>0$ and $9M^2 G^2 \Lambda<1$ \cite{GH} so that the above equation is integrated as,
\begin{equation}
t=r+\alpha \ln \left| \dfrac{r-r_H}{r_H}\right|+\beta \ln \left|\dfrac{r-r_C}{r_C}\right|+\gamma \ln \left|\dfrac{r-r_U}{r_U}\right|
\end{equation}
in which,
\begin{equation}
\alpha=\dfrac{3r_H}{\Lambda(r_C-r_H)(r_H-r_U)},\;\;\; \beta=\dfrac{-3r_C}{\Lambda(r_C-r_H)(r_C-r_U)},\;\;\; \gamma=\dfrac{-3r_U}{\Lambda(r_C-r_U)(r_H-r_U)}\nonumber
\end{equation}
and
\begin{equation}
r_H={\dfrac{2}{\sqrt{\Lambda}}}\cos\left[{\dfrac{1}{3}}\cos^{-1}(3M\sqrt{\Lambda})+\dfrac{\pi}{3}\right], r_C={\dfrac{2}{\sqrt{\Lambda}}}\cos\left[{\dfrac{1}{3}}\cos^{-1}(3M\sqrt{\Lambda})-\dfrac{\pi}{3}\right], 
r_U=-(r_H+r_C)\nonumber
\end{equation}
are the roots of the equation $dr/dt=0$. In fact, both $r_H$  and  $r_C$ have positive values, thereby defining the two horizons of the spacetime which are albeit of very different physical natures. The larger 
root $r_C$ is the de Sitter cosmological horizon but the smaller root `$r_H$' is the black hole event horizon. The negative root $r_U$ is not physical. Under the condition $3M\sqrt{\Lambda}<1$, the 
Schwarzschild black hole is sitting inside the cosmological horizon. Naturally, we should be able to obtain the Hawking radiation from either of the horizons. Here, we only consider the case of the 
emission from the event horizon, and later in the next section will come back to that of the cosmological horizon.
Utilizing the same approach employed in the previous sections, we consider an outgoing radial light ray which propagates from a point very close to inner horizon $(r_{in}=r_H+\epsilon)$ at $t=t_{in}$ to the 
detecting event ${\cal P}(t,r)$ where $r_H \ll  r\ll r_C$ and $ \epsilon \ll  r_H$, for an observer Bob at the finite radial coordinate $r$ from the event horizon. With the required initial condition, we will 
find the corresponding trajectory in the following form,
\begin{equation}
r \simeq t-t_{in} + \alpha\; {\ln} \;\dfrac{\epsilon}{r_H}
\end{equation}
So, the redshifted frequency `$\Omega$' will be related to the initial frequency at `$r=r_H+\epsilon$' by, 
\begin{equation}
\Omega \cong \Omega_{in} \;\exp(-\frac{t-t_{in}-r}{2 \alpha})
\end{equation}
Similar to the computations before, we find the asymptotic frequency spectrum to be,
\begin{equation}
|f(\omega)|^2 \varpropto \left[\exp\dfrac{12\pi r_H\omega}{\Lambda (r_C-r_H)(r_H-r_U)} -1\right]^{-1}
\end{equation}
which in terms of the energy modes becomes a Planckian power spectrum at the temperature, 
\begin{equation}\label{41}
T={\dfrac{\hbar}{2\pi}}\dfrac{\Lambda(r_C-r_H)(r_H-r_U)}{6r_H}
\end{equation} 
This again coincides with the known result obtained  through the conventional approach \cite{GH}. 
In the case of the Schwarzschild-dS spacetime, one may also consider to apply the same approach to the modes which are initially emanated from the cosmological horizon. 
This will be postponed to the next section when we consider Hawking radiation from the observer-dependent horizons.
%%%%%%%%%%%%%%%%%%%%%%%%%%%%%%%%%%%%%%%%%%
\subsection{Schwarzschild-AdS Horizons}
In this section, we work out the example of the Schwarzschild horizon with the AdS asymptotics (i.e $\Lambda < 0$ ). Given the metric \eqref{sch-ds}, the location of the event horizon in this case is 
determined by the zeros of the cubic equation,
\begin{equation}
f(r)=r^3+b^2r-2mb^2=0 \;\;\;;\;\;\; b^2 \equiv -\frac{3}{\Lambda}.
\end{equation}
As the function $f$ is strictly monotonic, this equation has only one root, given by \cite{SH,COV},
\begin{equation}
r_H={\dfrac{2b}{\sqrt{3}}}\;\sinh\left[{\dfrac{1}{3}}\sinh^{-1}(3\sqrt{3}\;\dfrac{m}{b})\right]
\end{equation}
Indeed, in this example a black hole horizon is always present. By expanding $r_H$ in terms of $m$, with $b \gg 3/m$, we have $r_+=2m(1-4m^2/b^2 + \cdot \cdot \cdot)$, which shows that the 
$\Lambda < 0$ has a shrinking effect on the horizon scale. For the outgoing light rays, the radial geodesics adopt the simple form,
\begin{equation}
\dfrac{dr}{dt}=1-\dfrac{2M}{r}+\dfrac{r^2}{b^2}\equiv {\dfrac{1}{b^2r}}(r-r_H)(r^2+rr_H+\rho^2)
\end{equation}
The solution is given by,
\begin{equation}
t={\dfrac{b^2r_H}{2r_H^2+\rho^2}}\left[{\ln}(\frac{r-r_H}{r_H})-\dfrac{1}{2} {\ln}(r^2+rr_H+\rho^2)+\gamma \tan^{-1}(\dfrac{2r+r_H}{\beta})\right]
\end{equation}
where $$\alpha={\rho^2}/r_H\;\;\;;\;\;\;\beta^2  = 4 \rho^2-r_H^2\;\;\;;\;\;\;\gamma=(2\alpha+r_H )/\beta\;\;\;;\;\;\;\rho^2=2mb^2/r_H$$ 
Considering a light ray propagating from $r_{in}=r_H+\epsilon$ at $t=t_{in}$ to the detecting event ${\cal P}(t,r)$ where $r_H \ll  r\ll b$ and $ \epsilon \ll  r_H$, the last equation becomes, 
\begin{equation}
t-t_{in}\simeq \;-{\dfrac{b^2\; r_H}{2r_H^2+\rho^2}} \;{\ln}\frac{\epsilon}{r_H}
\end{equation}
The redshifted frequency $\Omega$ will be related to the frequency at $r_{in}$ by,
\begin{equation}
\Omega \simeq  \Omega_{in}\left[\dfrac{(r_{in}-r_H)(r^2+rr_H+\rho^2)}{b^2r_H}\right]^{1/2} \simeq \Omega_{in}\exp\left[-{\dfrac{2r_H^2+\rho^2}{2 b^2r_H}}(t-t_{in})\right]
\end{equation}
As in the examples before, we find that the power spectrum for a wave packet scattered off the event horizon and traveled to infinity has the Planckian form at the temperature,
\begin{equation}
T={\dfrac{1}{4\pi}}\dfrac{2r_H^2+\rho^2}{b^2\;r_H} \end{equation}  
This again matches the result obtained in the conventional approach \cite{HP} and reduces to that of the asymptotically-flat Schwarzschild horizon for $b \rightarrow \infty$.
%%%%%%%%%%%%%%%%%%%%%%%%%%%%%%%%5 
\subsection{Cylindrical Horizons}
The third example that we probe in this section is the case of black holes with cylindrical event horizon and with the AdS asymptotics \cite{L}. By a coordinate transformation, 
the metric can be brought to the form,
\begin{equation}
ds^2=(b^2r^2-\dfrac{a}{br})dt^2-(b^2r^2-\dfrac{a}{br})^{-1}dr^2-b^2r^2dz^2-r^2d\phi^2\nonumber
\end{equation}
where $a \equiv 4M$ and $b^2 \equiv -\Lambda/3$. This is the metric of a static black hole whose event horizon is at $b r = a^{1/3}$, and has a curvature singularity at $r=0$. 
Similar to the examples before, we consider radially outgoing light rays which propagate from the proximity of the event horizon `$r_{in}=a^{1/3}/��b - \epsilon$, at $t=t_{in}$, 
to the detecting event ${\cal P}(t,r)$ where $r b \ll a^{1/3}$ and $ \epsilon \ll  r_+$. The equation governing the radial outgoing light rays in this space, can be written as follows,
\begin{equation}
\dfrac{dr}{dt}=-(b^2r^2-\dfrac{a}{br})
\end{equation}
By solving this equation (using relation 2.145 in \cite{GMR}), we obtain, 
\begin{equation}
t={\dfrac{1}{a^{1/3}b}}\left[{\dfrac{1}{6}}\; {\ln} \left(\dfrac{b^2r^2+bra^{1/3}+a^{2/3}}{(br-a^{1/3})^2}\right)+\dfrac{1}{\sqrt{3}}\tan^{-1}\left(\dfrac{2br+a^{1/3}}{-a^{1/3}\sqrt{3}}\right)\right]
\end{equation}
Introducing `$a^{\prime} = a^{1/3}$', with `$r\ll a^{\prime}$' and `$\epsilon\ll r_+$', the above relation becomes,
\begin{equation}
t-t_{in}\simeq -{\dfrac{1}{3ba^{\prime}}} \log \dfrac{b\epsilon}{a^{\prime}\sqrt{3}}
\end{equation}
The frequency $\Omega$ will be related to the initial frequency $\Omega_{in}$ at  $r_{in}$ by
\begin{equation}
\Omega (r\ll r_+) =  \Omega_{in}\left [\dfrac{br(b^3r_{in}^3-a^{1/3})}{br_{in}(b^3r^3-a^{1/3})}\right]^{1/2} \simeq \Omega_{in}\left(\dfrac{br}{a^{1/3}}\right)^{1/2}
\cdot\left(\dfrac{\sqrt{3}b\epsilon}{ba^{1/3}}\right)^{1/2} \simeq \Omega_{in}\exp\left[-\dfrac{3}{2}b{a^{1/3}}(t-t_{in})\right]
\end{equation}
Now if we repeat the analyses of the compact-horizon cases for the cylindrical horizon at hand, the asymptotic spectrum of the null rays will be thermally Planckian as follows,
\begin{equation}
|f(\omega)|^2 \varpropto \left[\exp\left(\dfrac{4\pi }{3b{a^{1/3}}}\right)\omega -1\right]^{-1}\;\;\;;\;\;\;T= \;\frac{3\hbar b a^{1/3}}{4\pi}
\end{equation} 
which is again the same result obtained through the canonical formalism \cite{L,13} .
%%%%%%%%%%%%%%%%%%%%%%%%%%%%%%%%%%%%%%%%%%%%5
\section{Hawking Radiation From The Observer-Based Horizons} 
A distinct class of causal screens, in both classical and quantum gravity, is the  category of \emph{observer-based horizons}. The Rindler horizon of an  accelerated observer in Minkowski spacetime is of course the 
canonical example of such horizons. A global spacetime can be deconstructed into, and in reverse 
re-assembled by, a sufficient number of such causal observer-based causal patches. It should be highlighted that, this is far from being a mere mathematical deconstruction of global spacetimes, because between these complementary 
causal patches there can be physical interplay. For example, the vacuum properties of the Minkowski spacetime is directly related with the maximal entanglements that interconnect the partnering 
pairs of the Rindler wedges. These `inter-regional maximal entanglements', once re-expressed in an underlying abstract quantum formulation of the geometry, can be proposed as the fundamental fabricators of the physical
spacetime \cite{26entnglmspt}.
Let us highlight that the idea of constructing an entire spacetime by reassembling a sufficiently many number of observer-based \emph {causal units} can serve us as a natural, and also
very productive approach in developing the complete theory of quantum gravity for generic spacetimes, including those which are phenomenologically significant for us. A very intriguing observer-based 
frame work of this kind, based on the identification of the \emph{causal diamonds} as the constituting causal units, has been already proposed and developed \cite{TBA}.\\ As the simplest model theory, one could begin 
with developing a microscopic theory of one Rindler patch \cite{24SR}. But a much more gratifying causal-unit theory would surely be the microscopic theory of the \emph{dS causal patch}, namely the 
static de Sitter patch proposed in \cite{BFA,TBA}. After obtaining the correct theory of static de Sitter patch, we could then look for the microscopic theory of the global de sitter spacetime. 
But, the correct microscopic theory of the dS causal patch is what we need phenomenologically.
With this perspective, through the rest of this section we will consider both of the aforementioned canonical observer-based horizons. The aim as in previous sections is to employ our approach to study the 
semiclassical behavior of the distantly propagated Hawking modes. This study will also clarify two more facts. First, as a by-product, it will independently reconfirm the important statement that, 
the observer-based causal horizons do behave in the same way in their semiclassically coarse-grained and  fine-grained aspects of the Hawking radiation as event horizons do. Secondly, there will be universal 
lessons on the mechanism of unitarity for the evaporation of observer-based causal screens that we take from the analysis here, to be utilized in the second  part of this study.
%%%%%%%%%%%%%%%%%%%%%%%%%%%%%%%%%%%%%%%%%%%%%%%%%%%%%%%%%%%5
\subsection{Hawking Modes From The de Sitter Cosmological Horizon}
W begin the examples of this section with the de Sitter cosmological horizon, while for a discussion of the comparative physics of the de Sitter spacetime in static and dynamic patches we refer to 
\cite{133233MNZJKHRA}. The static patch of the de Sitter spacetime is defined by the metric,
\begin{equation}
ds^2 = \left(1-{\dfrac{\Lambda}{3} r^2} \right)  dt^2 - \left(1-{\dfrac{\Lambda}{3} r^2}\right)^{-1} dr^2 - r^2d\Omega^2\nonumber
\end{equation}
The \emph{cosmological horizon} is the causally encompassing $\mathbb{S}_2$-surface with the finite radius $R = \sqrt{3/\Lambda}$. Therefore, the causal patch has also the finite radial extension 
$r \in (0,R)$. The late time observer Bob is located \emph{inside} the cosmological Horizon at a point sufficiently far from it. Employing our principal thought experiment, now consider an ingoing radial 
 ray which is emitted at $r_{in} = R - \epsilon$, with $\epsilon \ll R$, in a LIF in the proximity of the cosmological horizon at the initial time $t=t_{in}$, and so propagates toward the distantly located
Bob who much later receives those modes at the spacetime event $\mathcal{P}(r,t)$ with $r \ll R = \sqrt{3/\Lambda} \equiv H^{-1}$. The trajectory of the radial light ray can be written in the following  form,
\begin{equation} 
\dfrac{dr}{dt} =-\left(1-{\dfrac{\Lambda}{3} r^2}\right). 
\end{equation}
The integrated trajectory, given the required initial condition, is,
\begin{equation}
t-t_{in} = \dfrac{-1}{2H}\left(\log \dfrac{1+Hr}{1-Hr} - \log \dfrac{1+Hr_{in}}{1-H_{in}r}\right) \simeq \dfrac{-1}{2H} \log \dfrac{H\epsilon}{2}. 
\end{equation}  
So, the relation between the Bob's observed frequency and the initial frequency is given by,   
\begin{equation}
\Omega(r \ll H^{-1})=\Omega_{in} \left(\dfrac{1-H^2r_{in}^2}{1-H^2r^2}\right)^{1/2} \simeq \Omega_{in}\;(2H\epsilon)^{1/2} \simeq 2\; \Omega_{in}\;e^{-H(t-t_{in})}
\end{equation}
Now by repeating the same analysis we performed in the case of black hole horizons, we find in exactly the same way that the spectrum of the modes received by Bob follows the distribution, 
\begin{equation}
|f(\omega)|^2 \varpropto \left[\exp(\dfrac{2\pi \omega}{H}) -1\right]^{-1}
\end{equation}
which in terms of the energy quanta is a Planckian power spectrum at Hawking temperature,
\begin{equation}
T={\dfrac{\hbar}{2\pi}} \cdot \frac{1}{R}
\end{equation}
which is the result initially obtained in \cite{GH}.
As a follow up, let us go back to the example of the Schwarzschild-de Sitter spacetime and redo the analysis we did there, but now for a monochromatic light ray which propagates towards the interior 
observer from the near cosmological horizon. Trajectory of an ingoing null ray which starts from a point very close to the cosmological horizon $r_{in}=r_C-\epsilon$, takes the following form,
\begin{equation}
r \simeq -(t-t_{in}) + \beta\; {\ln}\dfrac{\epsilon}{r_H}
\end{equation}
And so, the frequency $\Omega$ will be related to the initial frequency $\Omega_{in}$ at $r_{in}$ by 
\begin{equation}
\Omega(r_H \ll  r\ll r_C) \simeq \Omega_{in} \left(\dfrac{2\epsilon}{r_C}\right)^{1/2} \simeq \Omega_{in}exp\left({\dfrac{t-t_{in}-r}{2\beta}}\right)
\end{equation}
Therefore, we will obtain a Planckian power spectrum defined at the temperature,
\begin{equation}
T={\dfrac{\hbar}{2\pi}}\;\dfrac{\Lambda(r_C-r_H)(r_C-r_U)}{6r_H}
\end{equation} 
which could be compared with the temperature given by \eqref{41}.
%%%%%%%%%%%%%%%%%%%%%%%%%%%%%%%%%%%%%%%%%%%5 
\subsection{Rindler Observer And Hawking Modes}
Now we come to our second canonical example, namely the Hawking radiation of the Rindler horizon. Besides the highlights we briefly pointed out in the beginning of this section, another important fact 
about Rindler horizon is that, for a large variety of phenomenologically interesting black holes, the near horizon region is given by the Rindler spacetime. Assuming that the microscopic theory 
of all the causal horizons should be the same, then this near-horizon similarity to Rindler may suggest one to postulate an intrinsic self-similarity in the underlying microscopic theory of black 
holes. This point will be discussed elsewhere \cite{HUBMM}.
Applying our approach to study the semiclassical spectral behavior of the Hawking modes for an evaporating Rindler horizon we 
consider a congruence of worldlines corresponding to a family of observers moving with constant acceleration $a$ in Minkowski spacetime. Parametrizing this family 
of worldlines in terms of the conventional coordinates $(\tau,\xi)$  \cite{GPB}, the metric takes the form, 
\begin{equation}
ds^2=dt^2-dx^2=(1+a\xi)^2 d\tau^2-d\xi^2\nonumber
\end{equation}
In fact, the proper coordinate system is incomplete and covers only a quarter of the Minkowski spacetime $x>|t|$, the so called \emph{Rindler wedge}. Indeed, this is the subdomain of Minkowski
spacetime which is causally accessible to a uniformly accelerated observer, namely, it is the causal patch of the accelerating observer, bounded by the corresponding observer-based causal horizon. 
The Rindler observer perceives this causal horizon at proper distance $a^{-1}$. Due to the presence of a causal horizon for the Rindler observer, one should expect a typical Hawking emission associated with it. 
To this end, we consider a radial light ray which propagates from $\xi_{in}=-a^{-1}+\epsilon$ at $\tau = \tau_{in}$ to an observer located in the spacetime position $(\xi,\tau)$, where $\xi\gg a^{-1}$ 
and $\epsilon \ll a^{-1}$. The trajectory of such radial light rays is given by the equation,
\begin{equation}
\dfrac{d\xi}{d\tau}=1+a\xi
\end{equation}
so that for the required initial condition the integrated trajectory will be,
\begin{equation}
\tau-\tau_{in}=\dfrac{1}{a} \log \left(\dfrac{1+a\xi}{1+a\xi_{in}}\right)\simeq \dfrac{1}{a} \log \left(\dfrac{\xi}{\epsilon}\right).
\end{equation}
Hence, the observed frequency $\Omega$ at $\xi\gg -a^{-1}+\epsilon$ will be related to the initial frequency $\Omega_{in}$ at $\xi_{in}=-a^{-1}+\epsilon$ by,
\begin{equation}
\Omega(\xi\gg -a^{-1})=\Omega_{in}\left(\dfrac{1+a\xi_{in}}{1+a\xi}\right)\simeq \Omega_{in}(\dfrac{\epsilon}{\xi})=\Omega_{in} e^{-a(\tau-\tau_{in})}
\end{equation}
Now, by a similar procedure employed in the previous cases, the Planckian spectrum for the observed Hawking modes is obtained to be, 
\begin{equation}
|f(\omega)|^2 \varpropto \left[\exp\left(\dfrac{2\pi \omega}{a}\right) -1\right]^{-1}
\end{equation}
corresponding to the Hawking temperature, 
\begin{equation}
T = \dfrac{\hbar a}{2\pi}
\end{equation}
Which is the standard Unruh temperature \cite{MSV}.
%%%%%%%%%%%%%%%%%%%%%%%%%%%%%%%%%%%%%%%%%%%%%%%%%%%%%%%%%%%%%%%%%%%%%%%%%%%%%%%%%5
%\newpage
\section{Unitarity, Holography and Horizons}
\subsection{The Unitarity of Information Processing In Bob's View : The Paradox} 
Suppose that an asymptotic observer \emph{Bob} is monitoring the entire history of a well-defined physical event in which a large black hole is formed out of a 
sufficiently high-energy initial scattering, 
or from the gravitational collapse of some sufficiently massive quantum matter. The so-formed black hole then evaporates, very softly but continually, by 
radiating-out quantum-sourced Hawking modes 
until finally annihilates into the asymptotic Hawking radiation, collectible by Bob. 
By the largeness of the black hole we mean that the radius of the initially-formed horizon, to be called \emph{the primal horizon}, is so enormous in Planck 
units that no quantum gravitational corrections would 
matter at the horizon scale (up until the horizon becomes of the Planck size, sufficiently-close to its total evaporation). To have the right setting 
for the information paradox, we also assume that the initial 
state of the quantum matter is set to be, and does stably remain to be, in \emph{a pure state}, before the primal horizon is being formed. That is, as 
an initial-state condition, we take the von Neumann entropy of the total quantum system in the \emph{pre-horizon era} to be zero,
\begin{equation} 
S^{{\rm\;Initial\;State}}_{({\rm von\;Neumann})} = 0 
\end{equation}
Now let us see, from Bob's information-theoretic view, the post-horizon-formation era. Bob's main assumption is that the evolution of the black hole, from the 
initial scattering or collapse to its total 
annihilation, is microscopically processed in a way that is completely `consistent with' all the structure and the laws of quantum physics. Indeed, given all his
experiences in string theory and quantum gravity, 
he does not have any serious evidence or even any significant motive to postulate any statement in contradiction to this assumption. on the other hand he does have 
a good number of evidences for this one 
assumption. For example, all black holes which are addressable by the AdS/CFT correspondence are confirmed to be so. Bob will so be assuming that this quantum 
gravitational phenomenon evolves in consistency with 
the unitary dynamics of a quantum system. \emph{A unitary dynamics preserves the information}, so, all the information that identify the exact initial state of 
the pre-horizon quantum system should be indestructible.
The delicate central question to be addressed is  \emph{whether or not unitarity of the information processing is validated by Bob}. To the asymptotic observations 
of Bob, the so-formed \emph{horizon} 
behaves by all means as a thermal physical membrane to which a thermodynamical entropy, the Bekenstein-Hawking entropy, equal to one quarter of its area in 
Planck units, is associated, 
\begin{equation}\label{bhe} 
S_{B.H} = \frac{A_{\;{\rm Horizon}}}{4 l_p^2} .
\end{equation}
since a quantum-finite positive entropy has developed in the gravitational system, Bob should be immediately concerned if the principle of information 
indestructibility is being held, due to 
the defining relation between entropy and information. Fortunately, by simply recollecting and utilizing the very defining virtue
of a black hole he gets the right clue out of this worry. A black hole is a maximally 
entropic gravitating system, entirely enclosed by a spacetime hypersurface, the horizon, that acts as a \emph{causal boundary} to all the events in the complementary part of 
the spacetime geometry. As such, the 
internal system of the black hole, which is localized inside the horizon, is causally disconnected from the asymptotic Bob, and so does not belong to 
his own causal patch. Now, merely by 
this defining criterion, Bob concludes that the indestructibility of the initial-state information becomes exactly equivalent to the following statement:\\ 
\emph{The Bekenstein-Hawking entropy (\ref{bhe}) associated to the primal horizon, now interpreted as the holographic statistical entropy of the horizon-interior 
system, should necessarily count 
the total number of black hole's microstates to be exactly equal to the total number of the initial-state information bits.} 
Let the integer $N$, which is taken to be a sufficiently large number, denote the total number of the initial-state information bits, 
\begin{equation} 
N \equiv I^{{\rm\;Initial\;State}}.
\end{equation}
Then, in Bob's view, the principle of information indestructibility is given by,
\begin{equation}\label{p1} 
S_{{\rm\; Primal\;Horizon}} = I^{{\rm\;Initial\;State}} 
\end{equation} 
\begin{equation}\label{p2} 
A_{\;{\rm\; Primal\;Horizon}}  = 4 l_p^2 \cdot N 
\end{equation}
As implied by the principal equalities (\ref{p1},\ref{p2}), Bob also concludes that, \emph{the initial-state information was built up, `bit by bit', the total Hilbert 
space of the primal black hole, by a one 
to one mapping that takes those bits to the black hole microstates}, that is,
\begin{equation} \rm{\;dim}[\mathcal{H}_{\;{\rm\;Primal\;Black\;Hole}}] = \rm{\;dim}[\mathcal{H}_{\;{\rm\;Primal\;Horizon}}] 
= \exp [\;I^{{\rm\;Initial\;State}}\;] =  e^N 
\end{equation}
But in fact as Bob experimentally confirms by his detector, the primal black hole is quantum unstable.,it decays 
and shrinks in size and mass, very very softly but monotonically, by a continual emission of Hawking modes. As the horizon decays such, the dimension of the black 
hole's Hilbert space, 
$\mathcal{H}^{\rm\;Interior}$, also shrinks in accordance with the Bekenstein-Hawking entropy formula. But then after this evaporation process is completed, once 
more the entirety of the
spacetime geometry becomes causally connected. Therefore, Bob will be able to check whether the initial-state information are all preserved, upon collecting the 
totality of the \emph{final-state} asymptotic Hawking 
modes. Now, to manifest the core of the problem, let us see \emph{how the dimension of $\mathcal{H}^{\rm\;Interior}_{(t)}$ at an arbitrary time $t$ after the 
formation of the primal horizon would shrink upon the 
release of a typical Hawking mode}. Also, to be more specific, let us consider the case of a Schwarzschild black hole in four spacetime dimensions in 
the rest of this study, knowing that an obviously 
similar analysis goes through in other spacetime dimensions, or for more complicated geometries. At  time $t$ of this quantum evaporation process, the horizon-interior 
system and the exterior spacetime geometry 
can be sufficiently-well described by those of a Schwarzschild geometry which is endowed with a very-slowly varying time-dependent mass $M(t)$, together with its 
corresponding quantities for the horizon 
radius $R(t)$, for the temperature $T(t)$, and for the black hole entropy $S_{{\rm Interior}}(t)$,  
\begin{equation} 
R(t) = 2 l_p^2\; M(t)\;\;\;;\;\;\;T(t) \equiv \beta^{-1}(t) = \frac{1}{8\pi l_p^2\;M(t)}\;\;\;;\;\;\;S_{{\rm Interior}}(t) = 4 \pi l_p^2\; M^2(t)  
\end{equation}
Now, let us suppose that at about this time $t$ a typical Hawking mode is emitted. The emission of this quanta reduces the entropy of the interior 
black hole system by the following amount,
\begin{equation}\label{deltas} 
\delta S_{{\rm\;Interior}}(t) = - \beta(t)\;\cdot\;\varepsilon_{{\rm\; H.M}}(t) 
\end{equation}
in which `$\varepsilon_{{\rm H.M}}(t)$' denotes the energy of the emitted Hawking mode at the emission time `\emph{t}. Having a thermal spectrum, the emitted 
Hawking mode is of the energy,
\begin{equation} 
\varepsilon_{{\rm H.M}}(t) = T(t) 
\end{equation}
and so from (\ref{deltas}) we learn that,
\begin{equation}\label{er1} 
\delta S_{{\rm Interior}}(t) = - 1.
\end{equation}
Namely, each typical Hawking mode, once being radiated, carries away one unit of entropy from the black hole. Combining this with the unitarity 
conditions (\ref{p1},\ref{p2}), we obtain, 
\begin{equation}\label{er123} 
S_{{\rm\;Interior}}(t) = N - N^\prime(t)
\end{equation} 
\begin{equation}\label{er321} 
{\rm dim}[\mathcal{H}_{\rm interior}(t)] = e^{N-N^\prime(t)} 
\end{equation} 
with $N^\prime(t)$ denoting the total number of the Hawking modes radiated-out up until a time `$t$' after the formation of the primal horizon. 
Of course we know that the emitted Hawking modes obey the Planckian spectrum, mostly carrying $O(1)$ entropy. 
However, doing a statistically more careful analysis 
with all details included, we would get the same result for $S_{{\rm\;Interior}}(t)$ as stated above. From (\ref{er123},\ref{er321}) Bob concludes that, 
\emph{the primal black hole will be all annihilated
after radiating out a total number of $N$ thermal Hawking quanta, namely as many as the total number of the initial-state information.} 
Now, let us suppose that Bob monitors the black hole for a long time scale which turns out to be of the order of $\tau \propto M^3 \propto N^{\frac{3}{2}}$ \cite{DP}, 
until the primal black hole evaporates entirely. 
If so, then the\emph{final state} of the system would be an asymptotic radiation characterized by a \emph{thermal 
density matrix $\hat{\rho}$} defined at the 
temperature, 
\begin{equation}\label{ton} T = (4 \sqrt{\pi} l_p)^{-1} N^{-\frac{1}{2}} 
\end{equation}
whose von Neumann entropy, $S \equiv - {\rm Tr}[\hat{\rho}\;\log(\hat{\rho})]$, is equal to,
\begin{equation} 
S_{{\rm Asymptotic\;Radiation}}^{\;({\rm Thermal\;State})} = N 
\end{equation} 
and whose expectation value of the total number-operator is also equal to the same $N$. Now, in such a quantum many body system, in which the von Neumann 
entropy saturates its
upper bound in being equal to the total number of the microscopic degrees of freedom of the
system, the ﬁnal state would be information-free, namely
\begin{equation} 
I ^{{\rm\; Final\;State}} = 0, 
\end{equation}
which means that all the initial-state information, in a maximum violation of unitarity, is lost. This `quantum information mismatch' 
does clearly call Bob for a resolution.
%%%%%%%%%%%%%%%%%%%%%%%%%%%%%%%%%%%%%%%%%%%%%% 
\subsection{Geometric Planckianity and the optimal disperser: The Microscopic Theory}
To resolve this quantum information mismatch, which violates unitarity as maximally as possible, one should begin with pinpointing the exact 
origin of the (almost-) Planckianity 
of the asymptotic Hawking radiation, as the very first step.
In the first part of this study, we approached the above issue in an \emph{observer-centric} method
whose generality and stability was verified by scanning over a broad class of causally-nontrivial spacetimes that collectively constituted a landscape 
of all the major deformations. The general physical theorem, which we validated by probing broad examples, is the following statement:\\ 
\emph{An initial-state system of the near-horizon single-frequency Hawking modes becomes a final-state system of  
Hawking modes with Planckian spectrum, merely as the direct result of 
the geodesic propagation of the frequency modes through the curved background of the asymptotic observer's causal patch.}
Now, to further advance this understanding, we should first appreciate that, this maximal dispersion of the frequency modes is equivalent to the 
quantum-information mismatch itself. As we found there, 
this maximal dispersion of the frequency modes is a characteristically classical geometric effect which 
disturbs a microscopic quantum system. 
Having known the basic points, we are finally at the right stage here to pinpoint the correct origin of the information paradox. The final picture will become 
transparent in several steps. Let us first 
ask the following question.\\
What is the underlying microscopic mechanism of this conversion of the initial system of the near-horizon single-frequency modes into a final system of the 
 asymptotic modes with Planckian spectrum?\\
In answering this question, this much of the microscopic mechanism is known  that operating on the quantum system of frequency modes, and maximally disperses them, 
it should be the classical limit of 
a \emph{geometric environment}. In a better wording, the frequency modes constitute \emph{an open quantum system} which interact with the quantum environment of 
some characteristically-geometric degrees 
of freedom. The collective outcome of this environment-system entanglement, in a certain classical limit, is the geometric effect by which the 
modes are optimally 
dispersed. We call this optimally-dispersing environment, the \emph{optimal disperser}. Now as the next step forward, we should develop the structure of 
an associated Hilbert space scheme which 
microscopically formulates the above mentioned physical picture. \emph{First}, at any arbitrary time during the horizon evaporation, there is one Hilbert space accounting 
for the totality of 
the emitted Hawking modes denoted by $\mathcal{H}_{{\rm\; Hawking\;Modes}}^{(t)}$. This Hilbert space, representing an evolving open quantum system, should be  accommodated 
into a total Hilbert space defining an entire closed quantum system. Let us highlight that we are defining this total Hilbert space for \emph{the smallest set of the microscopic degrees of freedom 
which can be consistently treated as a closed system in Hawking evaporation}. So,
\begin{equation}\label{tfhm}
\mathcal{H}_{{\rm\; Hawking\;Modes}}^{(t)} \;\subset_{{\otimes}_{(t)}}\; \mathcal{H}_{{\rm\;Total}}^{({\rm Minimum})} 
\end{equation}
As our $\mathcal{H}_{{\rm\; Hawking\;Modes}}^{(t)}$ defines all the Hawking modes available in the \emph{Bob's causal patch} up until a time $t$ after the formation 
of the primal-horizon, 
its dimension is increasing as far as the black hole evaporation continues. Therefore, in (\ref{tfhm}), the left-hand side is explicitly \emph{time-dependent}, but 
the right-hand side is static. 
Next by(\ref{tfhm}), we identify the subsystem \emph{complementary} to the Hawking Modes, 
\begin{equation}\label{tfcmpl} 
\mathcal{H}^{\rm \;Complementary}_{(t)}  \;\subset_{{\otimes}_{(t)}}\; \mathcal{H}_{{\rm\; Total}}^{({\rm Minimum})} 
\end{equation} 
and in accordance with the physical description given before, it has to develop the optimum dispersion in a specific environmental limit.
As a marked note, let us clarify a point of fundamental importance. By thinking more fundamentally one may begin the formulation with $\mathcal{H}_{{\rm\;Total}}$ 
which  defines the Hilbert space of
the total microscopic system of this scenario to be \emph{any, or a smallest choice, of the Bob's causal diamonds which accommodates the entire 
history of the evolution of the black hole}, from the initial high-energy scattering or collapse to the completion of its Hawking emission \cite{BFhorizon,HUBMM}. 
Because a causal diamond should be 
physically treated as a consistent closed universe \cite{TBA}, the Hilbert space of Bob's causal diamond defines the \emph{Theory of Everything}, the \emph{TOE}, 
for his causal universe 
in which $\mathcal{H}_{{\rm\; Total}}^{({\rm Minimum})}$ is set \emph{as a sub-theory}. 
In what follows we will restrict our analysis to the $\mathcal{H}_{{\rm\; Total}}^{({\rm Minimum})}$-theory, and refer the interested readers to consult the 
forthcoming study on Bob's unified theory 
\cite{HUBMM}. Let us simply restate that in such a theory, the relation (\ref{tfhm}) will be replaced by,
\begin{equation} 
\mathcal{H}_{{\rm\;Total}}^{({\rm Minimum})} \;\subset_{\otimes}\;\mathcal{H}^{{\rm\;Bob}}_{({\rm TOE})} 
\end{equation}
Now, let us go back to the defining relation (\ref{tfcmpl}), and ask the following important question:\\ 
\emph{What is the physical identity of the microscopic
degrees of freedom defined by $\mathcal{H}^{{\rm \;Complementary}}_{(t)}$?} 
To answer this question, we need to understand more clearly the fine-grained,
microscopic nature of the classical geometric mechanism which, in the coarse-grained collective way of the first part, disperses the initial system
of single-frequency Hawking modes into a final system of Planckian modes. It is expected that the underlying fine-grained structure is inversely 
obtainable by \emph{deconstructing} the 
background curvature of Bob's causal diamond with a null-horizon causal boundary down to a Hilbert space of quantum curvature modes which build up that 
geometric environment. 
Being translated and promoted into an exact statement about the Hilbert space systems defined by the relations (\ref{tfhm},\ref{tfcmpl}), we are led to the 
following principal identification:\\
\emph{$\mathcal{H}^{{\rm \;Complementary}}_{(t)}$, defined by relations (\ref{tfhm},\ref{tfcmpl}) at any time $t$ of Hawking evaporation process, should be  
identified as the Hilbert space 
of the  microscopic degrees of freedom' that build up the curvature of the asymptotic-observer's  causal diamond with a null-horizon 
causal boundary, at the corresponding time $t$ . That is,}
\begin{equation}\label{mgeons} 
\mathcal{H}^{{\rm \;Complementary}}_{(t)} \equiv \mathcal{H}^{[{\rm Causal\;Diamond}]}_{{\rm\;Bobby\;Curvature\;Modes}\;(t)} 
\equiv \mathcal{H}^{{\rm\;Bobby\;Causal\;Curvature}}_{(t)} 
\end{equation}    
According to this principal identification, we call the degrees of freedom of $\mathcal{H}^{{\rm\;Bobby\;Causal\;Curvature}}_{(t)}$ as the Bobby curvature modes. 
From the above identification together with the relations (\ref{tfcmpl}) and (\ref{tfhm}), we learn that our total Hilbert space admits the following 
tensor decomposition,as an exact equality which should hold at any time during the entire Hawking evaporation, 
\begin{equation}\label{tfeq1} \mathcal{H}_{{\rm\;Total}}^{({\rm Minimum})} = 
\mathcal{H}^{{\rm\;Bobby\;Causal\;Curvature}}_{(t)}\;\otimes\;\mathcal{H}_{{\rm\; Hawking\;Modes}}^{(t)} 
\end{equation}
In the rest of this paper, we will be referring to the quantum many body system which is defined by the triplet system
of `$( \mathcal{H}_{{\rm\; Hawking\;Modes}}^{(t)},\mathcal{H}^{{\rm\;Bobby\;Causal\;Curvature}}_{(t)},\mathcal{H}_{{\rm\;Total}}^{({\rm Minimum})} )$' 
the \emph{Hilbert space triplet}, or simply \emph{the quantum triplet}.
Introducing the physical identification (\ref{mgeons}), and the Hilbert space structure (\ref{tfeq1}), now we should revisit the optimum dispersion 
demand which we introduced earlier, 
and examine carefully its defining physical conditions. Let us first restate the demand as follows:
\begin{multline}\label{cfl} \;{\rm\;Optimal\;disperser\;:}\;{\rm \;\exists\;A\;Specific\;Semiclassical\;Limit\;of\;The\;Quantum\;Triplet\;s.t:}
\hspace*{1.55 cm}\\\hspace{3.5 cm}{\rm\;Optimal\;disperser} 
= \left(\mathcal{H}^{{\rm\;Bobby\;Causal\;Curvature}}_{(t)}\right)_{{\rm\;In\;That\;Specific\;Semiclassical\;Limit}}  
\end{multline}
So then we ask: What physical conditions do identify the optimum dispersion Limit as phrased in the demand (\ref{cfl})? 
The question, in better wording, is the following:
What are the necessary and sufficient conditions to be imposed on the quantum many body system defined by our Hilbert space triplet so
that the optimum dispersion limit of 
the demand (\ref{cfl}) is realized?\\ 
To answer this question, first we note that the Planckian dispersion of the propagating modes requires as a necessary condition that the Bob's causal geometry
should be curved in the presence of a null causal screen. Therefore, by implementing this basic understanding into the exact microscopic physics of the Hilbert space triplet, we now state the first
necessary condition for the realization of the specific semiclassical limit that corresponds to the \emph{optimal disperser} as follows:\\
\emph{C.1}. The semiclassical optimum dispersion limit can be realized by the triplet
$( \mathcal{H}_{{\rm\; Hawking\;Modes}}^{(t)},\mathcal{H}^{{\rm\;Bobby\;Causal\;Curvature}}_{(t)},\mathcal{H}_{{\rm\;Total}}^{({\rm Minimum})} )$, if,  as the first necessary condition, 
this triplet corresponds to a quantum many body system that consistently defines the curved geometry of the sufficiently large causal diamond of an
asymptotic observer which has as its  causal boundary the null horizon. 
But, this necessary condition is not yet a sufficient one. Namely, based on the major lessons we took from the analyses of the first part of this study, 
we still need to impose one more condition 
 on the triplet system for the optimum dispersion to be realized by it. This second necessary condition will be the direct quantum microscopic 
 implementation of the following statement into the Hilbert space triplet:\\
\emph{The optimal disperser of the frequency modes of an open quantum system, can be actualized only in the strict 
thermodynamic limit of the quantum environment defined and constituted by the Bobby curvature modes.}\\
Here, the thermodynamic limit, or the continuum limit, is meant exactly as in the (classical or quantum) statistical physics of many body systems, namely the exact limit in
which the total number of the microscopic degrees of freedom of the system goes to \emph{infinity}. Accordingly, the second necessary condition for the microscopic 
realization of the optimal disperser limit is stated as follows:\\
\emph{C.2}. The optimal disperser can be realized by the triplet quantum system (\ref{tfeq1}) only in the strict thermodynamic limit of the environment 
of the Bobby Curvature Modes,
\begin{equation}\label{ojc} 
\;\;\;{\rm dim}[\;\mathcal{H}^{{\rm\;Bobby\;Causal\;Curvature}}_{({\rm\;Optimal\;disperser\;Limit})}\;]\;=\;\infty\;\;\; 
\end{equation} 
Therefore, for the Bobby-bulk geometry of a spacetime possessing a quantum-evaporating horizon to disperse the initial-state information to the Planckian effect (with or without the 
grey-body factors or all the other perturbative corrections included in it), \emph{both of the conditions \emph{C1} and \emph{C2} should be simultaneously realized by the quantum triplet}. But a
careful examination reveals the following fact:\\ 
\emph{Fact:\; Once the Hilbert space triplet $( \mathcal{H}_{{\rm\; Hawking\;Modes}}^{(t)},\mathcal{H}^{{\rm\;Bobby\;Causal\;Curvature}}_{(t)},\mathcal{H}_{{\rm\;Total}}^{({\rm Minimum})} )$ 
satisfies the condition \emph{C1}, then the condition \emph{C2} can not be realized}.\\
One can see that a microscopic validation of \emph{C1} invalidates \emph{C2} by simply applying the universal statements of holography \cite{HSBHR123M}, as well the Jacobson's 
rederivation of the Einstein's equations \cite{Jacobson}, to the Bobby-causal type configurations as phrased in the statement of \emph{C1}. It is being concluded that, for the quantum many 
body system defined by the Hilbert space triplet to realize the condition \emph{C1}, it should satisfy the following \emph{holographic demand on the dimension of the total quantum system}, 
\begin{equation} \label{hd}  
{\rm dim}[\mathcal{H}_{{\rm\;primal\;Horizon}}] 
= \exp(\frac{A_{\;{\rm Primal\;Horizon}}}{4 l_p^2}) = e^N = {\rm dim}[\;\mathcal{H}_{{\rm\;Total}}^{({\rm Minimum})}\;] 
\end{equation} 
\emph{Now}, although the right hand side of the \emph{`holographic equality'} (\ref{hd}) can be arbitrarily large, it has to be a \emph{finite} integer. 
Indeed, the area of the primal horizon in Planck 
units can be taken as large as one wishes, but it should always be \emph{finitely large}, simply because the strict limit of an \emph{infinite horizon radius} 
is physically 
inconsistent with the presence of an asymptotically far observer `Bob' to collect and analyze the asymptotic Hawking radiation. Because of this strict finiteness, 
and by utilizing the equalities (\ref{tfcmpl}) and (\ref{mgeons}) which identify $\mathcal{H}^{{\rm\;Bobby\;Causal\;Curvature}}$ as a subspace 
of $\mathcal{H}_{{\rm\;Total}}^{({\rm Minimum})}$, we conclude that, 
\begin{equation}\label{cnoj} {\rm dim}[\;\mathcal{H}^{{\rm\;Bobby\;Causal\;Curvature}}_{(t)}\;] \;<\; \infty\;\;\;;\;\;\;\forall t 
\end{equation} 
This does contradict (\ref{ojc}) which is the statement of the necessary condition \emph{C2}. This proves that a simultaneous validation of the two necessary 
conditions \emph{C1} and \emph{C2} of an optimal
disperser environment for the radiated Hawking modes,  is physically impossible. Therefore we have learned the following lesson:\\ 
\emph{A holographically correct $\mathcal{H}^{{\rm\;Bobby\;Causal\;Curvature}}$ can never be an optimal disperser. Therefore, the unitarity of the information processing 
in the Hawking evaporation of
the causal horizons is guaranteed if `holography' is implemented microscopically.}\\
It is interesting to highlight one independent point. We have been explicit in the condition C2 on the statement that, the optimal disperser can only be realized if the 
Bobby curvature environment 
is infinitely large in its microscopic degrees of freedom. This point may deserve a distinct attention. By naturally taking this statement beyond its specific quantum 
gravity context into quantum information 
theory, it becomes an information theoretic conjecture which we do propose.
This information theoretic conjecture' is the abstract general statement that, the Planckian-effect dispersion of 
any set of frequency modes in any subsystem of a given total closed system can not happen as far as the total system is finite dimensional.
By the term Planckian-effect dispersion here, we collectively mean any \emph{environmentally sourced} dispersion of frequency modes which is of the Planckian type, 
incorporating the possible grey-body
factors and any possible type of perturbative quantum corrections in it. We should also note that this information theoretic statement is consistent with the Page 
phenomenon which is generic to the
finitely-large quantum many body systems \cite{Page}.
Now, let us get back to the central subject of this section, and restate our main conclusion. In short, we have learned that, the optimum dispersion can only be 
realized in a limit of Bob's 
causal-curvature environment that is holographically illicit. Therefore, there can never be any information loss in holographic microscopic systems.
%%%%%%%%%%%%%%%%%%%%%%%%%%%%%%%%%%%%%%%%%%%%%%%%%%%
\subsection{The Holographic Way To Quantum Gravitational Unitarity}
Finally, we are at the right stage to clarify in more details how \emph{holography} should be implemented into the observer-centric  microscopic theory of  
the triplet quantum many body system  
$( \mathcal{H}_{{\rm\; Hawking\;Modes}}^{(t)},\mathcal{H}^{{\rm\;Bobby\;Causal\;Curvature}}_{(t)},\mathcal{H}_{{\rm\;Total}}^{({\rm Minimum})} )$ which formulates the 
\emph{minimum total system} of a quantum-evolving causal horizon during its entire history.\\ 
Naturally anticipated, it becomes manifest that the correct microscopic theory of the Hawking-evaporating horizons as described in the asymptotic observer's causal patch is being formulated 
based on an intrinsically-holographic quantum duality between the horizon-interior system and the causally-independent system of Bobby curvature modes. That, in 
one way or another, the quantum 
systems in the \emph{interior} and the \emph{exterior} of a causal horizon should be physically \emph{mirroring} one another is a major idea in quantum gravity 
that dates back to the seminal papers which 
proposed the principle of \emph{observer complementarity} \cite{MSV}. Indeed, the quantum duality known as observer complementarity is rooted in the black hole 
information physics.
It was shown that for consistently joining the natural postulate of the quantum unitarity of the Hawking evaporation process with 
the equivalence principle, one should also postulate a 
\emph{quantum duality} between a defining pair of 
Hilbert spaces in the interior and in the exterior of the causal horizon. However, many physical and mathematical details of the corresponding \emph{duality map} 
have been a matter of both minor and major
revisits up until present \cite{MSV}, which is yet to be discovered. Here we state and utilize our distinct version of this inside-outside duality. 
The duality that we frame here will also 
be at the level of an exact equivalence between a pair of Hilbert spaces in the two causal sides of the horizon, but the detailed duality map between the degrees 
of freedom on both sides will be an interesting 
theme for future works. The exact foundation of this inside-outside duality proposal will be unfolded from first principles in the forthcoming paper \cite{HUBMM}. 
Here, we will briefly elaborate on the reasoning behind this proposal.
Let us remember that $\mathcal{H}^{{\rm\;Bobby\;Causal\;Curvature}}_{(t)}$ is by definition the quantum system of all the Bobby curvature modes at the given 
time, namely it is the system of all
those degrees of freedom that microscopically build up the time-$t$ \emph{curvature} of the causal patch of an asymptotic observer Bob who is monitoring the entire 
evolution of the horizon.
Let us also remember that, the evolving horizon monotonically shrinks by the slow radiation of Hawking modes into the Bob's causal patch, so that the black hole's 
mass continually decreases up until it vanishes
entirely. Accordingly the curvature of the Bob's causal geometry, which is directly sourced by the decreasing mass of the black hole, is also being driven to 
vanish monotonically. For example,
in our canonical model, the bulk spacetime curvature is simply modeled by that of the Schwarzschild metric with a monotonically-decreasing time-dependent mass 
profile is it. Now, it is holographically clear
that the total number of the Bobby curvature modes should be counted by the area in Planck units of the causal horizon, monotonically decreasing from its 
initial value $N$ to zero. Therefore, let us 
introduce a one-parameter family of \emph{integers} that serve as a holographic measure of the time-dependent area of the evaporating horizon in Planck units,  
\begin{equation} 
N(t) \equiv \frac{{\rm\;Horizon\;Area(t)}}{4 l_p^2} \;\;\;;\;\;\; N(t) \in \mathbb{N} \;\;\;;\;\;\;\forall t  
\end{equation}
Now, the dimension of the Hilbert space of the curvature modes is given by,
\begin{equation}\label{eql92} 
{\rm dim}[\;\mathcal{H}^{{\rm\;Bobby\;Causal\;Curvature}}_{(t)}\;] = e^{N(t)}\;\;\;\;\;\;\forall t
\end{equation} 
which evolves in accordance with the following boundary conditions on $N(t)$, 
\begin{equation}\label{bvs} N(0) = N = I_{\rm\;Initial\;State}  \;\;\;;\;\;\;N(t > t^\star) = 0 \end{equation}
with $t^\star$ denoting the time when the evaporation is completed.
Indeed, on the account of the to-be proposed \emph{inside-outside duality}, there is another Hilbert space, which although is defined independently of
the triplet $( \mathcal{H}_{{\rm\; Hawking\;Modes}}^{(t)},\mathcal{H}^{{\rm\;Bobby\;Causal\;Curvature}}_{(t)},\mathcal{H}_{{\rm\;Total}}^{({\rm Minimum})} )$, 
and does not belong to it directly, has 
the exact same dimension at any time during the entire evolution of the horizon. Obviously, by the first statement of holography \cite{HSBHR123M}, this 
independent Hilbert space is that 
which defines the microstates of the black hole in the interior of its causal horizon. That is, one has this all-time valid holographic equality 
of dimensions,
\begin{equation}  
 {\rm dim}[\;\mathcal{H}_{{\rm\;Horizon\;Interior}}^{(t)}\;] = e^{N(t)} = {\rm dim}[\;\mathcal{H}^{{\rm\;Bobby\;Causal\;Curvature}}_{(t)}\;] \;\;\;;\;\;\; \forall t  
\end{equation}
This holographic equality of dimensions for two independently-defined Hilbert spaces which belong to a pair of causal-complementary regions of the same global 
spacetime, signals that there may 
be also a much deeper physical connection between them. This anticipation turns out to be indeed credential, and all that it takes to unfold its physical 
statement is to pinpoint yet another instrumental 
role played by \emph{holography} for the physics of horizons. The central holographic point is that we have a setting in which the same holographic 
screen \cite{HSBHR123M} is shared 
by the two quantum theories defined by $\mathcal{H}_{{\rm\;Horizon\;Interior}}^{(t)}$ and $\mathcal{H}_{{\rm\;Bobby\;Causal\;Curvature}}^{(t)}$. 
On one hand, the quantum system of black hole microstates living on $\mathcal{H}_{{\rm\;Horizon\;Interior}}^{(t)}$ possesses as its 
holographic screen the causal boundary 
of the horizon interior, which is identical with the horizon surface. On the other hand, \emph{Bob's TOE} which includes the complete 
quantum gravitational physics of his causal patch is also an intrinsically holographic theory \cite{TBA}. Namely, Bob's TOE which lives 
on $\mathcal{H}^{{\rm\;Bob}}_{({\rm TOE})}$  and 
is defined on the basis of holographic screen of any sufficiently-large causal diamond that accommodates the entire evolution of the horizon. 
As Bob's holographic screen, whose area in Planck units should 
be at least of the order of $N^{\frac{3}{2}}$, will be enormously larger that the primal horizon whose holographic size is $N$, we may initially think 
that it should have nothing to do with 
the primal horizon which is the holographic screen for the the theory of black hole microstates. In fact the correct
physics is going to be fruitfully different \cite{HUBMM} and it will be shown that the basic conclusion is the following statement,\\ 
\emph{The two holographic quantum theories living on $\mathcal{H}_{{\rm\;Horizon\;Interior}}^{(t)}$  and $\mathcal{H}^{{\rm \;Bobby\;Causal\;Curvature}}_{(t)}$, 
being a pair of causally-independent quantum many body systems which nevertheless share the same system-defining holographic screen, are quantum dual to 
one another, and so their exact Hilbert space physics could 
be  mapped to one another. That is, the following statement of inside-outside duality does hold,}
\begin{equation}\label{hologradual} 
\;\;\;\mathcal{H}_{{\rm\;Dual\;Horizon\;Interior}}^{(t)} \equiv \mathcal{H}^{{\rm \;Bobby\;Causal\;Curvature}}_{(t)} \;\;\;;\;\;\;\forall t\;\;\; 
\end{equation}
Based on the above duality, the internal degrees of freedom that microscopically construct the time-dependent system of the black hole in the horizon interior,
and the quantum curvature modes in the causal patch of the asymptotic observer Bob, are mapped to one another  and are reconstructible from each similar to 
what has been shown in \cite{MSV}. One point to be
highlighted here is that, this pair of systems in the inside and in the outside are always holographic subsets of the total microscopic degrees of 
freedom that define the primal horizon. 
Now, what about the open quantum system of the Hawking modes living in the Bob's bulk? Because the two quantum systems living 
on $\mathcal{H}_{{\rm\;Hawking\;Modes}}^{(t)}$ and 
$\mathcal{H}^{{\rm \;Bobby\;Causal\;Curvature}}_{(t)}$ are complementary pairs of our quantum triplet whose dimension is fixed by (\ref{hd}), and given that the 
time-dependent dimension of $\mathcal{H}^{{\rm \;Bobby\;Causal\;Curvature}}_{(t)}$ is set by the equality (\ref{eql92}), we learn that the total dimension of 
the Hawking-mode subsystem evolves as follows, 
\begin{multline}\label{hmd} 
\hspace{3.55 cm} {\rm dim}[\;\mathcal{H}_{{\rm\;Hawking\;Modes}}^{(t)}\;] = \exp [N - N(t)] \equiv e^{N^{\prime}(t)} \\ {\rm\;With\;The\;Boundary\;Conditions}:\; \\ 
N^{\prime}(0) = 0\;\;\;;\;\;\;N^{\prime}(t > t^\star) = N =  I_{{\rm\;Initial\;State}} \hspace{3.5 cm}
\end{multline}
According to the time-dependent dimension (\ref{hmd}), as the horizon evaporates, the subsystem of the emitted Hawking modes expands in population from 
zero to a total number of $I^{{\rm\;Initial\;State}}$
once the horizon becomes entirely annihilated. In other words in the process of horizon evaporation, 
the initial \emph{Bobby curvature modes} $I_{{\rm\;Initial\;State}}$  will be converted into the asymptotic $I_{{\rm\;Initial\;State}}$ Hawking modes as 
the final state of the system. The extremely
important point is that, as was holographically proved in the previous section,
\emph{during this holographic conversion of the modes the initial-state bits of information are all indestructibly preserved, and so are being unitarily 
processed as the horizon evaporates}.
Because of this holographic unitarity, the final state of this evolving system will be  a \emph{pure quantum state}, that is, in terms of its von Neumann 
entropy it is characterized by the 
following condition,
\begin{multline}\label{pc}\hspace{5.85 cm} S^{{\rm\;Final\;State}}_{({\rm von Neumann})} = 0\\\ I^{{\rm\;Initial\;State}} = I^{{\rm\;Initial\;State}} \hspace{5.25 cm}
\end{multline}
Next, let us describe in some more detail how this purity of the final state of the Hilbert space triplet is processed in the form of a quantum many body system 
of the $N = I^{{\rm\;Initial\;State}}$ microscopic modes  live in the Bobby causal bulk.
To this end we describe how the projected distribution of the microscopic modes evolve in different patches. The Hilbert space
triplet $( \mathcal{H}_{{\rm\;Hawking\;Modes}}^{(t)},\mathcal{H}^{{\rm\;Bobby\;Causal\;Curvature}}_{(t)},\mathcal{H}_{{\rm\;Total}}^{({\rm Minimum})})$
identifies a total number of $N$ microscopic degrees of freedom in the Bob's causal patch which are distributed among its
interacting, simultaneously evolving, complementary quantum subsystems $\mathcal{H}^{{\rm\;Bobby\;Causal\;Curvature}}_{(t)}$ and $\mathcal{H}_{{\rm\;Hawking\;Modes}}^{(t)}$ in
a unitarily-driven $\left(N(t),N^\prime(t)=N-N(t)\right)$-profile. The point is that the (holographically projected) localization of these $N$ degrees of
freedom is done in a way that characteristically distinguishes between the modes distributed in the two complementary subsystems. Let us see this characteristic difference. 
The $\mathcal{H}^{{\rm\;Bobby\;Causal\;Curvature}}_{(t)}$-subsystem is constituted from the modes that by definition build up the curvature of the Bob's causal patch.
So, simply by their physical identification, it is clear that those modes should have a radially inhomogeneous distribution in the Bob's bulk. 
That the localization of the $N(t)$ Bobby curvature modes should be regionally marked with a radially-decaying profile of the mode-density is 
manifested by the fact that the curvature-invariants of the spacetime  are radially enhanced by 
nearing the horizon. The most familiar example of this behavior is the Kretschmann invariant of Schwarzschild spacetime with a $1/r^{6}$ profile. Moreover, 
the distribution of the Bobby curvature modes should be dominantly concentrated in a non-asymptotic subregion near the horizon which we will discuss below. 
Not only the rapidly decreasing power-law profile of the curvature invariants shows this near-horizon dominance of 
the Bobby curvature modes, but also we will see a reaffirmation of this fact in the discussion below on the semiclassical finite-distance spectrum of the Hawking modes.
In contrast, the $\mathcal{H}_{{\rm\; Hawking\;Modes}}^{(t)}$-subsystem is composed of the $N^\prime(t) = N -N(t)$ radiated Hawking modes which are dynamically 
classified into 
two sub-groups of modes with \emph{complementary patch localization, and also distinct physical behavior}. To see this inter-grouping of Hawking modes in 
our observer-centric approach, we should go back to the principal analysis of the first part, but this time redo it for an observer who is located in an arbitrarily 
finite distance from the horizon. By the geometric mechanism with which these modes become semiclassically dispersed while propagating in the observer's bulk, the 
following fact is clear. Given a spacetime with a null causal horizon, there should be a finite crossover scale $R_C$, so that the emitted Hawking modes which 
propagate to distances  beyond this characteristic scale are observed by the local observer Bob to be (semiclassically) dispersed to a Planckian spectrum (up to 
unimportant finite-size corrections). In a single-scale  Schwarzschild spacetime, $R_C$ can only depend on the horizon radius, and therefore it should be simply 
proportional to it, namely, $R_C \propto R_H \propto N^{\frac{1}{2}}$. But even in multi-scale spacetimes, it is still obvious that, the most dominant dependence
of $R_C$ should be carried by the horizon scale in the same linear manner. Moreover, being mainly a result of the extremely dominant role that 
the exponential redshift virtue of the null horizon plays in the semiclassical dispersion, we expect that (modulo some sub-leading dependences),
$R_C \sim a \cdot R_H \sim a \cdot l_p\;N^{\frac{1}{2}}$, with $a$ being a constant, typically of order-one. To fix $a$, we will need the microscopic 
theory, but its value is not important to us. Let us name the horizon region $R \lesssim R_C$ the non-asymptotic region (NAR), and the region beyond it,
\emph{the asymptotic region}. Now, the fact that the semiclassical dispersion of Hawking modes is being developed already in the NAR, does reconfirm that the 
Bobby curvature modes are also concentrated in the NAR. So, we conclude that the dominant distribution patch of the modes of the quantum triplet is of the form,
\begin{multline} 
\hspace*{3.5 cm} \mathcal{H}^{{\rm\;Bobby\;Causal\;Curvature}}_{(t)}\; \approxeq\; \mathcal{H}^{{\rm\;NAR\;Curvature\;Modes}}_{(t)}\\
\mathcal{H}_{{\rm\;Hawking\;Modes}}^{(t)}\;  \approxeq\; \mathcal{H}_{{\rm\;NAR\;Hawking\;Modes}}^{(t)} \otimes \mathcal{H} _{{\rm\; Asymptotic\;Hawking\;Modes}}^{(t)}  
\hspace{2.1 cm}
\end{multline}
The quantum triplet system begins its unitary evolution in the \emph{pre-Page time} during which an enormous environment of the NAR curvature modes interacts with the NAR
Hawking modes. Then at a time scale of about the \emph{the Page time} \cite{Page}, the system turns to its \emph{post-Page time} era during which the NAR curvature 
modes become a small system 
of interacting defects among the radiated Hawking modes in the NAR. The system evolves unitarily, until  a time scale of the order $N^{\frac{3}{2}}$, it
ends up with $N = I^{{\rm Initial\;State}}$ asymptotic Hawking modes at Bob's disposal, defining a pure final state. Recollecting and joining all the stated points so 
far, we obtain the following structure for the triplet Bobby quantum many body system that holographically formulates the unitary evolution of the quantum-evaporating 
causal screens,
\begin{multline}\label{99} 
\hspace*{1.66 cm} \hspace*{.33 cm} \mathcal{H}_{{\rm\;Minimum\;Total}}^{({\rm dim} = e^N) }\;  \;=\; \mathcal{H}_{{\rm\;Dual\;Horizon\;Interior}}^{({\rm dim} = e^{N(t)})} \otimes_{(t)} 
\mathcal{H} _{{\rm\; Hawking\;Modes}}^{({\rm dim} = e^{N - N(t)})}\\ \approxeq\; \mathcal{H}_{{\rm\;NAR\;Curvature\;Modes}}^{({\rm dim} = e^{N(t)})} \otimes_{(t)} 
\mathcal{H} _{{\rm\; NAR\;Hawking\;Modes}}^{({\rm dim} = e^{N - N(t) - N_R(t)})} \otimes_{(t)} \mathcal{H} _{{\rm\; Asymptotic\;Hawking\;Modes}}^{({\rm dim} = e^{N_R(t)})} \hspace*{.55 cm} 
\end{multline}
This separation of scales which is seen in the holographic projection of the microscopic modes into the Bob's bulk should have a very meaningful impact  on the
quantum many body system of the primal horizon, which calls for further study \cite{HUBMM}. 
%%%%%%%%%%%%%%%%%%%%%%%%%%%%%%%%%%%%%%%%%
\section{A Brief Conclusion: An Observer-Centric Causal Holography} 
By utilizing an entirely observer-centric methodology, we have shown  how, for generic black holes and causal screens beyond the AdS/CFT family 
of such objects, 
the Hawking evaporation is indeed protected against any loss of the information by the virtue of a must-be-held holography which is defined in the causal patch of 
a sufficiently far observer. 
Indeed, in this same sense, the best route towards the ultimate and precise resolution of the information paradox turns out to be an \emph{entirely 
observer-centric causal holography} 
which is to be implanted in the very root-foundation of Quantum Gravity.Based on this approach, we have explored and detailed the microscopic physics of an 
observer-centric holography that does realize the 
unitarity of information processing in the quantum evolution of generic causal horizons.
However, the corresponding \emph{triplet quantum many body system}, summarized in (\ref{99}), clearly needs to be improved to become a complete microscopic 
description. In particular,
still there are three main aspects of this microscopic system yet to be known. 
First, the detailed microscopic interactions between all the different sets of modes needs to be fixed. 
Second, the exact unitary operator that drives the dynamics of the entire system, together with the individual dynamics of each one of the evolving subsystems 
should be determined. 
Third, we still need to unfold how the purity of the final state of the system is imprinted on the fine-grained  quantum correlations of the Hawking modes.
As a hint forward, by the holographic procedure in which Bobby curvature modes are being dynamically converted into the Hawking modes, and also by their holographic 
origin as the 
identical microscopic degrees of freedom of the many body system of the primal horizon as the Holographic screen, it is very natural to anticipate that, in the triplet quantum 
system there should be an underlying scheme of impartial physical unification between all those sets of modes. Namely, the fundamental microscopic description of the 
system should be such that all the modes are being physically considered as identical entities. Moreover, both from this study, and from several 
independent quantum gravitational reasonings, it becomes clear that this fundamental holographic microscopic formulation should be observer-centric in its very conception
and formulation. 
%%%%%%%%%%%%%%%%%%%%%%%%%%%%%%%%%%%%%%%%%%%%%%%%%%%%%%%%%%%%%%%%%
\section *{Acknowledgments}
Javad Koohbor and Mohammad Nouri-Zonoz would like to thank University of Tehran for supporting this research project under the grants provided by
the research council. Mohammad Nouri-Zonoz also thanks the Albert Einstein Center for Fundamental Physics, University of Bern, for kind hospitality 
and supporting his visit during which part of this study was carried out. Alireza Tavanfar thankfully acknowledges the support of his research by the Brazilian Ministry of Science, Technology and Innovation (MCTI-Brazil) and also in continuation from Funda\c{c}$\tilde{a}$o para a Ci$\hat{e}$ncia e a Tecnologia (Portugal) through the project UID/EEA/50008/2013. Alireza Tavanfar would also like to thank CERN Physics Department, where some of the ideas in the second  part of this research were initially developed, for both hospitality and support. 
%\pagebreak
%%%%%%%%%%%%%%%%%%%%%%%%%%%%%%%%%%%%%%%%%%%%%%%%%%%%%%5
%\\\\\hspace*{7 cm}{\Large References}
 \end{document}